\documentclass[journal]{IEEEtran}
\usepackage{ifpdf}

\usepackage{marginnote}

\usepackage{multirow}

\usepackage{cite}

\usepackage{xcolor}
\usepackage{mdframed}

\ifCLASSINFOpdf
  \usepackage[pdftex]{graphicx}
  \graphicspath{{./Images/Experiments/}{./Images/Introduction/}{./Images/MatrixPencil/}{./Images/Simulation/Distributed/}{./Images/Simulation/PointNew/}{./Images/Simulation/PointNew_1/}{./Images/Systemmodel/}{./Images/Experiments/Chimney/}{./Images/Experiments/ChimneyNew/}{./Images/Experiments/rain/}{./Images/Simulation/DistributedNew/}{./Images/Simulation/DistributedNew_1/}}
\else
   \usepackage[dvips]{graphicx}
\fi
\usepackage{amsmath}
\ifCLASSOPTIONcompsoc
    \usepackage[caption=false,font=normalsize,labelfont=sf,textfont=sf]{subfig}
\else
    \usepackage[caption=false,font=footnotesize]{subfig}
\fi
\usepackage{dblfloatfix}
\begin{document}
%
\title{Matrix-Pencil Approach-Based Interference Mitigation for FMCW Radar Systems}
%
%
%

\author{Jianping~Wang,~\IEEEmembership{Member,~IEEE,}
        Min~Ding,
        and~Alexander~Yarovoy,~\IEEEmembership{Fellow,~IEEE}
\thanks{The authros are with the Faculty of Electrical Engineering, Mathematics and Computer Science (EEMCS), Delft University of Technology, Delft, 2628CD the Netherlands. e-mail: J.Wang-4@tudelft.nl, min.dingchina@hotmail.com, A.Yarovoy@tudelft.nl.}
}

%
%

\markboth{Submitted to IEEE Transactions on Microwave Theory and Techniques}%
{Shell \MakeLowercase{\textit{et al.}}: Bare Demo of IEEEtran.cls for IEEE Journals}
%



\maketitle

\begin{abstract}
A novel matrix pencil-based interference mitigation approach for FMCW radars is proposed in this paper. The interference-contaminated segment of the beat signal is firstly cut out and then the signal samples in the cut-out region are reconstructed by modeling the beat signal as a sum of complex exponentials and using the matrix pencil method to estimate their parameters. The efficiency of the proposed approach for the interference with different parameters (i.e. interference duration, signal-to-noise ratio (SNR), and different target scenarios) is investigated by means of numerical simulations. The proposed interference mitigation approach is intensively verified on experimental data. Comparisons of the proposed approach with the zeroing and other beat-frequency interpolation techniques are presented. The results indicate the broad applicability and superiority of the proposed approach, especially in low SNR and long interference duration situations.
\end{abstract}


%
\IEEEpeerreviewmaketitle

\section{Introduction}
%
%
%
%
\IEEEPARstart{F}{requency} modulated continuous-wave (FMCW) radars are widely used in both civilian and military applications due to its simple processing method, high accuracy and high reliability. With the explosive increase of wireless radio and sensing applications, FMCW radars face increasingly severe interference from other devices. For instance, modern cars are equipped with multiple FMCW radars to assist drivers and improve transportation safety, where the radars inevitably cause strong interference among each other. Moreover, FMCW weather radars also suffer from the radio frequency interference from the surrounding environment. In these situations, the strong interference leads to reduced radar sensitivity and resolution, weak target masking and probably ghost target detection. Therefore, to overcome these problems and alleviate performance degradation of the radar systems, it is crucial to take proper interference mitigation in practice.

So far, a number of approaches have been proposed for interference migration, which can be mainly classified into two categories: (i) system-level approaches; (ii) post-signal processing techniques. System-level approaches exploits temporal, spatial, polarization, frequency and code diversities in radar system, antenna array and waveform design. In \cite{Kim2005}, a circular polarized antenna architecture is design to combat the linear polarized interference. Meanwhile, the frequency hopping technique learned from bats is also generally used to counteract various interference caused by spectrum congestion \cite{Bechter2016}. To identify mutual interference, the predefined orthogonal patterns \cite{Kim2016} are imposed on the frequency modulation slopes of each FMCW burst which consists of hundreds of sweeps. Medium Access Control (MAC)-like approach is proposed to regulate transmission time of the multiple radars in the same area \cite{Aydogdu2019,Khoury2016}. These approaches provide effective solution to interference mitigation, but they increase the complexity of radar system or antenna design for implementation and lead to costly systems.      

On the other hand, the post-signal processing techniques utilize a range of digital signal processing approaches to mitigate interference probably at the expense of increased computational load. The signal processing methods can be further divided into three classes: filtering approaches \cite{Choi2016,Feng2018}, signals separation \cite{Uysal2019,Ren2019}, and suppression and reconstruction approaches \cite{Tullsson1997,Neemat2019,Toth2019RadarConf}. In \cite{Choi2016}, weighted-envelope normalization approaches are proposed to deal with strong spiky mutual interference by detecting the envelope variations within a sliding time window and inversely normalizing the detected interference. In \cite{Feng2018}, an adaptive noise canceller is devised for mutual interference suppression by exploiting the different distributions of frequency spectra of target's signals and mutual interference in the frequency domain. However, both filtering approaches are only applicable to tackle certain type of interference or point-like targets scenario, which limits their wide applications. Meanwhile, the stability of the adaptive filter is hard to guarantee. 

The signals separation methods generally exploit different features, i.e., distinct sparsity of targets' signals and the interference in different transform domains to separate them \cite{Uysal2019,Ren2019}. So these methods require some prior information about the sparsity of the desired signal and the related interference to construct proper bases for optimal separation. However, if the ``off-grid'' problem between the bases (e.g., the discrete Fourier basis and short-time Fourier transform basis \cite{Uysal2019}) and the signal to be represented exists, it would lead to some loss of the degree of sparsity, thus degrading the separation performance. 

By contrast, as long as the extension of the interference is limited in a certain domain, the simplest but effective method to suppress the interference is, in practice, to directly cut the interference-contaminated samples out of the signal with various windows (e.g., zeroing and inverse cosine window)\cite{Babur2009,Babur2010}. However, the interference cutting-out not just eliminates the interference but also suppresses part of the useful signal of targets, which reduces the signal to noise ratio (SNR) of the targets after coherent processing and decreases the range resolution. To deal with the SNR loss problem, a Burg method-based interpolation was used to extrapolate the useful signal samples in the cut-out region in the time-frequency ($t$-$f$) domain \cite{Neemat2019}. It uses the signal samples on both sides of the cut-out gap to separately extrapolate the cut-out data forward and backward. Then, the forward- and backward-extrapolated samples in the cut-out region are summed up with weights by a specifically designed cross-fading window. This method is generally applicable to mitigate various interference for FMCW radars (as indicated in Fig.~\ref{fig:interference_4_cases} later). But its extrapolation accuracy degrades dramatically when the number of the cut-out samples of signals increases. In \cite{Toth2019RadarConf}, the signal extrapolation with AR model was suggested using the instrumental variable method (IVM). However, this method is not very stable and cannot always get proper signal reconstruction.                

\begin{figure}[!t]
	\centering
	\includegraphics[width=0.46\textwidth]{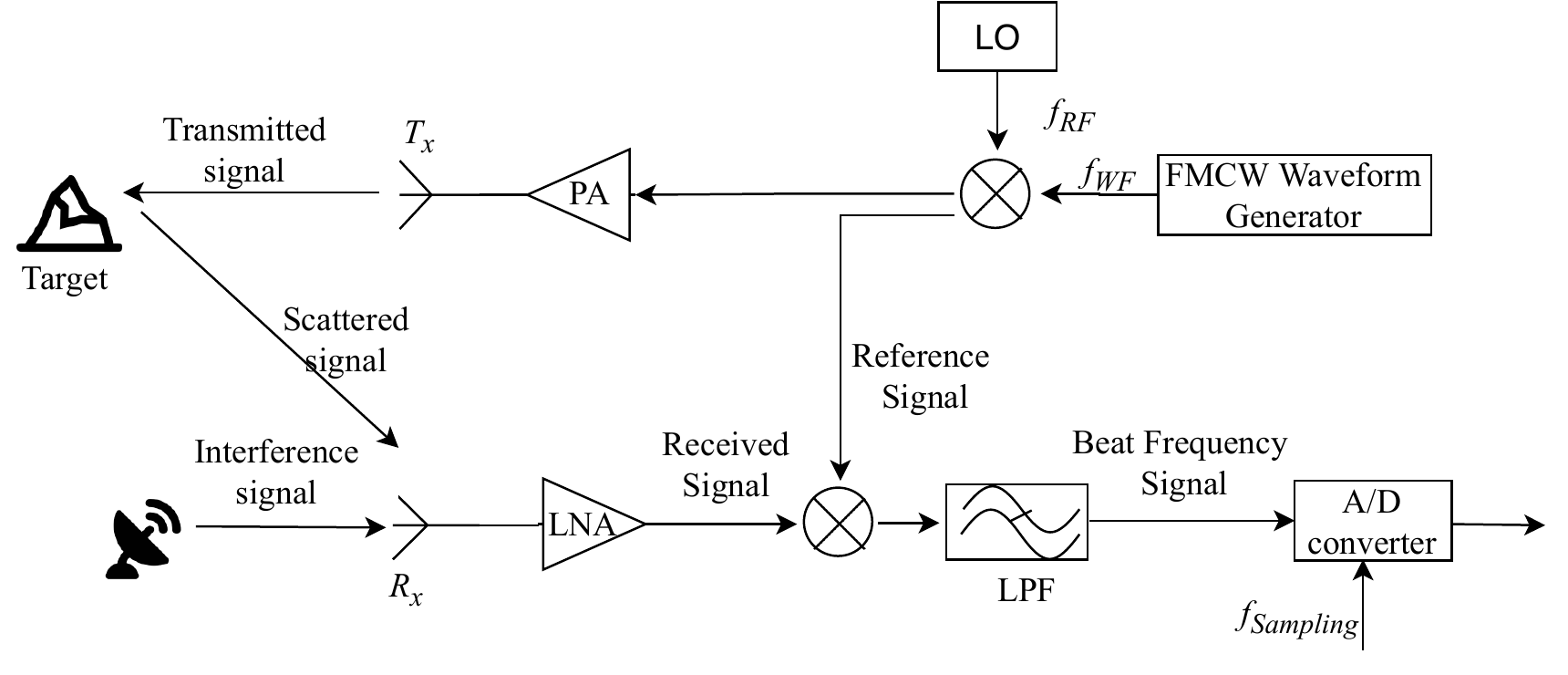}
	\caption{General block diagram of mono-static linear FMCW radar system.}
	\label{fig:FMCWPrinciple}
\end{figure}

To {accurately extrapolate the cut-out data after cut-out operation (i.e., zeroing)}, we propose an iterative matrix-pencil (MP) method-based extrapolation for interference mitigation. Similar to the Burg method-based approach, the proposed approach first cuts the interference-contaminated samples out of the signals and then reconstruct/extrapolate the clipped samples of the useful signals. But the proposed approach simultaneously accounts for the signals before and after the clipped samples by using a unified all-pole model which is derived from the analytical model of the beat signals of targets. So it provides the potential to get more accurate extrapolation of the non-contaminated signal in the cut-out region. Before the extrapolation, the all-pole model is first estimated based on the interference-free samples with the matrix-pencil method \cite{Sarkar1995,Hua1990}. However, in practice, the noise and the possible discontinuity of the interference-free samples would impact the accuracy of the estimated signal model, thus resulting in less accurate reconstruction of the cut-out samples of useful signals. To alleviate this effect, an iterative scheme is introduced to refine the model estimation and the extrapolation, which significantly improves the accuracy of the signals in the cut-out region. Moreover, we want to mention that a method similar to the one presented in this paper has been used for multi-band signal fusion for high-resolution imaging in \cite{Zou2016,Wang2018}. Actually, for interference mitigation, the measured signals become two or more separate segments after interference suppression. So, using the interference-free signal segments to reconstruct/extrapolate the cut-out region is in essence a signal fusion problem. The main difference is absence of the incoherence-correction between different signal segments needed for interference mitigation. Note this paper focuses on interference mitigation on sweeps in the time domain which would be flexible to be followed by other further processing. Nevertheless, we should mention that in the case of interference mitigation followed by some specific two dimensional (2-D) processing (e.g., range-Doppler processing, range-DOA estimation), the proposed interference mitigation approach could also be extended and implemented in the high-dimensional space by exploiting the 2-D or high-dimensional MP approaches \cite{Hua1992TSP,Chen2007TSP}, which would be considered in future.

The rest of this paper is organized as follows. Section~\ref{sec: FMCW_radar_system_model} formulates the basic models of the signals received by FMCW radars. In Section~\ref{sec: MP_based_Interf_mitig}, the proposed iterative matrix-pencil method based interference mitigation approach is presented. Then, its performance of interference mitigation is demonstrated in different scenarios through the numerical simulations in section~\ref{sec: Num_Simu} and the experimental results in section~\ref{sec: Exp_Results}. Finally, conclusions are drawn in section~\ref{sec: Conclusion}.

\section{FMCW Radar System Model}  \label{sec: FMCW_radar_system_model}

\subsection{Transmitted and received signals}
The system diagram of an FMCW radar system
is shown in Fig.~\ref{fig:FMCWPrinciple}. The transmitted FMCW signal can be expressed as 
\begin{equation}
	p(t) = A_{tx} \exp\left[j2\pi \left(f_0+\frac{1}{2}K t\right)t\right],
\end{equation}
for $0< t < T/2 $, where $A_{tx}$ is the amplitude of the transmitted signal, and $f_0$ is the starting frequency of an FMCW sweep. $K = B/T$ is the chirp rate defined by the ratio of the signal bandwidth $B$ and the sweep time $T$. The transmitted electromagnetic (EM) signal is intercepted by targets and scattered back to the receiver. Considering the quasi-monostatic configuration of the transmit and receive antennas and assuming single scattering process for each target, the back-scattered signal can be represented as 
\begin{equation} \label{Eq: RF receiver signal}
	s_r(t) =\sum_{i=1}^{M} A_{rx,i} \exp\left[j2\pi \left(f_0(t-t_i)+\frac{K}{2}(t-t_i)^2\right)\right] 
\end{equation}
where $t_i=2d_i/c$ is round-trip time delay of the scattered signal related to the $i^\text{th}$ target at a distance of $d_i$, and $A_{rx,i}$ is the corresponding amplitude of the signal which subsumes the scattering coefficient and propagation loss. $c$ is the speed of light and $M$ is the number of targets.


\subsection{Dechirp on receiver}

\begin{figure*}[!t] 
	\normalsize
	\begin{equation} \label{eq: signal_dechirp_filter}
	\begin{aligned}
	\Tilde{s}(t)
	&= \mathcal{F}_{lp}\{[s_r(t) + s_\text{int}(t)] \cdot p^\ast (t)\} \\
	&= \mathcal{F}_{lp}\big( s_\text{int}(t) \cdot p^\ast (t) \big) + \mathcal{F}_{lp} \left\{\sum_{i=1}^{M} A_{tx} A_{rx,i} \exp\left[-j2\pi \left( f_0 t_i - \frac{K t_i^2}{2} \right) \right] \cdot \exp \left(-j2\pi K t_i t \right) \right\}   \\
	&=\mathcal{F}_{lp} \big(s_\text{int}(t) \cdot p^\ast(t) \big) + \sum_{i=1}^{M^\prime} \Tilde{A}_{r,i}\exp\left[-j2\pi\left(f_0 t_i - \frac{K t_i^2}{2} \right)\right]\exp\left(-j2\pi K t_i t \right) 
	\end{aligned}
	\end{equation}
	\hrulefill
\end{figure*}

In FMCW radar system, dechirp processing is commonly used due to its simple operation and low requirement of sampling rate for the Analog to Digital Converter (ADC). It is implemented by mixing the received signals with the conjugate of the transmitted one, which leads to beat signals. 

Considering the occurrence of strong interference $s_\text{int}$, the beat signal after demodulating and filtering can be formulated as \eqref{eq: signal_dechirp_filter} on the top of next page, where the superscript~$^\ast$ denotes complex conjugate and $\mathcal{F}_{lp}$ is the low-pass filter operator. $\Tilde{A}_{r,i}$ is the amplitude of the received signal of the $i^\text{th}$ target and $M^\prime (\leq M)$ is the number of observed scatterers within the desired unambiguous range. As $\exp\left[-j2\pi \left( f_0 t_i - \frac{K t_i^2}{2} \right) \right]$ is a constant phase term related to the $i^\text{th}$ target which can be subsumed by the amplitude of the signal, one can present $a_i = \Tilde{A}_{r,i}\exp\left[-j2\pi \left(f_0 t_i - \frac{K t_i^2}{2} \right) \right]$ as a new complex signal amplitude. Then, \eqref{eq: signal_dechirp_filter} can be rewritten as a sum of complex exponential functions
\begin{equation}\label{Eq: Target beat signal} 
\Tilde{s}(t) = \mathcal{F}_{lp} \big( s_\text{int}(t) \cdot p^\ast (t) \big) + \sum_{i=1}^{M^\prime} a_i \exp\left( -j2\pi f_{b,i}t \right) 
\end{equation}
where $f_{b,i} = K t_i$ is the beat frequency corresponding to the $i^\text{th}$ target. For moving targets, $t_i=2d_i/c = 2(d_{i0}+v_i t)/c$ can be used to account for the Doppler shift, where $v_i$ and $d_{i0}$ are the velocity and the initial distance of the $i^\text{th}$ target relative to the radar. Generally, as $v_i \ll c$, it has negligible impact on the target's beat frequency within a short FMCW sweep. After getting beat frequencies, the ranges of different targets can be calculated as
\begin{equation}
   d_i = \frac{c\cdot f_{b,i} }{2K}
\end{equation}
As thermal noise and measurement errors always exist due to physical limitation of the practical radar system, the signal measurements can be modeled as
\begin{equation} \label{Eq: signal model in time domain}
\begin{aligned}
    s(t) &= \Tilde{s}(t) + n(t)\\
    &= \sum_{i=1}^{M^\prime} a_i \exp(-j2\pi f_{b,i} t) + \mathcal{F}_{lp}\big( s_{int}(t)\cdot p^\ast(t) \big) +n(t) \\
    &=\Tilde{s}_\text{tar}(t) + \Tilde{s}_\text{int}(t) +n(t) 
    \end{aligned}
\end{equation}
where $s(t)$ represents the measured signal, $n(t)$ denotes the noise and measurement errors,
$\Tilde{s}_\text{int}(t)=\mathcal{F}_{lp}\big( s_\text{int}(t) p^\ast(t) \big)$ is the signal resulting from the interference, and $\Tilde{s}_\text{tar}(t)=\sum_{i=1}^{M^\prime} a_i \exp(-j2\pi f_{b,i}t )$ is the beat signal of targets within the desired detection range. Equation \eqref{Eq: signal model in time domain} gives the general model of the FMCW radar measurements contaminated by strong interference.     

\subsection{Interference}

\begin{figure}[!t]
	\centering
	\includegraphics[width=0.49\textwidth]{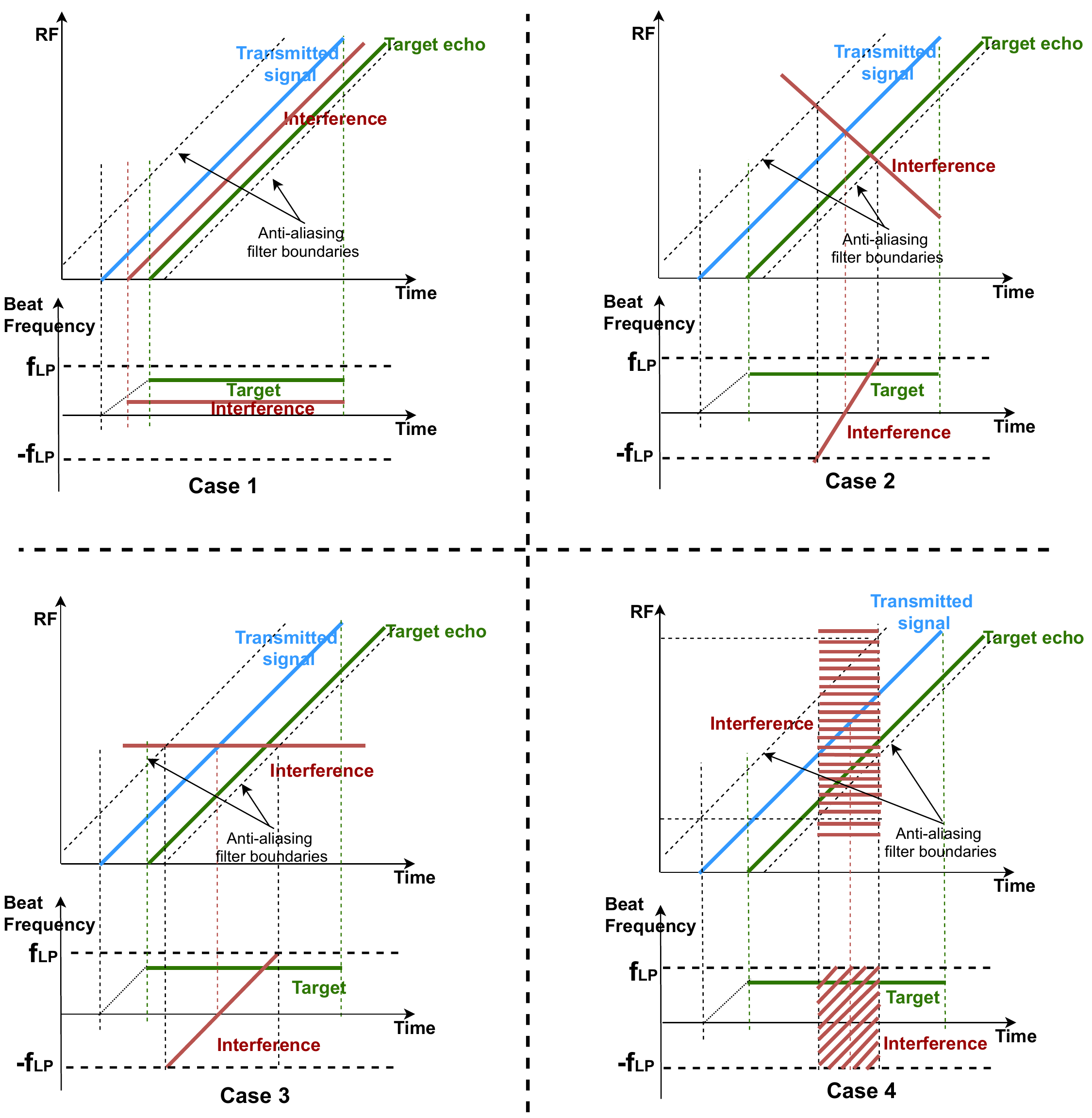}
	\caption{Four cases of interference which corrupt the FMCW radar system. Case 1: chirp interference with the identical sweep parameters as the victim radar; Case 2: chirp interference with different sweep parameters from the victim radar; Case 3: sinusoidal/narrowband continuous interference; and Case 4: instantaneous wideband interference.}  
	\label{fig:interference_4_cases}
\end{figure}

Nowadays, radar systems face various types of interference due to the rapid increase of radio wireless applications. In particular, for FMCW radar systems, the related interference can be classified as the following four cases \cite{Brooker2007,Brooker2010,Schipper2014}: 1) FMCW interference with the same chirp rate; 2) FMCW interference with a different chirp rate; 3) CW interference; and 4) transient interference. These cases are illustrated in Fig.~\ref{fig:interference_4_cases}. In Case 1), the FMCW interference would result in a strong ghost target if it appears within the reception window of the system determined by the maximum detection range. In Cases 2) and 3), the FMCW and CW interference have a long time duration and lead to the non-constant beat frequency after the dechirp processing. Thanks to the low-pass filtering, their occurrences are confined in a short time around the frequency intersecting moment. In Case 4), the spectrum of the transient (or pulse) interference with a rectangular amplitude in a short time can be considered as equidistant lines with a $\sin(x)/x$ envelope. Some of these frequency lines intersect with the reference FMCW signal of dechirp operation and then, as in Case 3), result in the short interference   after low-pass filtering \cite{Brooker2010}.

The above analysis indicates that the interference in Cases 2), 3) and 4) all cause contaminated measurements in certain time period within an FMCW sweep duration, which in principle can be tackled using the method described in this paper (Note the interference with a very small sweep slope difference from that of the victim radar (i.e., extreme situations in case 2) could make all the signal samples contaminated, in which case the proposed approach and other zeroing plus reconstruction methods would not be applicable). Without loss of generality, we consider the FMCW signal was contaminated by an FMCW interference with a different frequency slope, i.e., Case 2) in the following sections.

Assuming an interfering FMCW radar is located at a distance $d_I$ away from the transceiver, the interference signal arriving at the receiving antenna can be expressed as 
\begin{equation} \label{Eq: RF Interference signal}
    s_\text{int}(t) = A_I \exp \left [j2\pi \left( f_{I,0}(t-t_I)+\frac{K_I}{2}(t-t_I)^2 \right) \right]
\end{equation}
for $t_I < t < T_I + t_I $, where $A_I$ is the amplitude of the interference. $t_I=d_I/c$ is the time delay of the interference signal relative to the starting time of the transmission of the victim radar. $f_{I,0}$ is the starting frequency of the interference signal and $K_I = B_I/T_I$ is the chirp rate of the interference signal with  the bandwidth $B_I$ and the sweep duration $T_I$. 

Then, the interference signal $\Tilde{s}_\text{int}(t)$ obtained after dechirping and low-pass filtering can be explicitly expressed as
\begin{equation} \label{Eq: beat_signal_interference_LP}
\Tilde{s}_\text{int}(t) = \mathcal{F}_{lp}\left( s_\text{int}(t)p^\ast(t) \right) =\mathcal{F}_{lp} \left\{ a_I \exp \left[j \Phi(t)\right] \right\}
\end{equation}
where 
\begin{align}
    \Phi(t) &= 2\pi \left[ \left(\frac{K_I}{2} - \frac{K}{2} \right) t^2 + \left( f_{I,0} - f_0 - K_I t_I \right) t  \right] \\
    a_I &= A_I A_{tx} \exp \left[j 2\pi \left( \frac{K_I}{2}t_I^2 - f_{I,0} t_I \right)\right]
\end{align}

Taking the first derivative of the phase $\Phi(t)$ with respect to time, one can get the instantaneous beat frequency
\begin{equation} \label{Eq: beat_freq_interference}
    f_{b,I}(t)  = -\frac{1}{2\pi}\frac{\partial \Phi_I (t)}{\partial t} 
    = (K_1 t + K_2)
\end{equation}
where $K_1 = (K-K_I) $ and $K_2 = \left(f_0-f_{I,0} + K_I t_I\right)$ are constant coefficients. According to \eqref{Eq: beat_freq_interference}, the beat frequencies resulting from the interference are time-varying. After the low-pass filtering in \eqref{Eq: beat_signal_interference_LP}, its frequency bandwidth and the time of occurrence are confined but the time-varying property is not affected. By contrast, the beat frequencies of targets are constant, as shown in \eqref{Eq: signal model in time domain}.  This difference between the beat frequencies of targets and interferer makes the interference mitigation can be done in either time or time-frequency ($t$-$f$) domain \cite{Babur2009}.

\section{Matrix Pencil Method Based Interference Mitigation} \label{sec: MP_based_Interf_mitig}
A model-based interference mitigation approach for the FMCW radar system is presented in this section. This approach can operate in either the time domain or the time-frequency domain. Without loss of generality, its details are illustrated through the time-domain processing for the interference mitigation in the following sections.

\subsection{Discrete signal in the time domain}

From \eqref{Eq: signal model in time domain}, the discrete signal measurements can be written as
\begin{align} \label{Eq: discrete_signal_timeDom}
    s[k] &= \Tilde{s}_{\text{tar}}[k] + \Tilde{s}_{\text{int}}[k] + n[k] \nonumber\\
    &=\sum_{i=1}^{M^\prime} a_i z_i^k + \Tilde{s}_{\text{int}}[k] + n[k]
\end{align}
where $z_i = \exp{(j2\pi f_{b,i} \Delta t)}$, $\Delta t$ is the sampling interval and $k = 0,\ 1, ..., N-1$ is the sampling indices of the $N$ time-domain samples in an FMCW sweep. As analyzed above, the interference component $\Tilde{s}_{\text{int}}$ appears in a short period in a sweep; thus, only some of the measured signal samples, e.g. from $N_1$ to $N_2$ are contaminated, where $0\leq N_1 <N_2\leq N-1$. Since the desired targets' signal $\Tilde{s}_{\text{tar}}$ is a sum of exponential components, it is natural to suppress the interference by cutting out the contaminated samples from the measurements and then reconstructing the cut-out samples with the uncontaminated measurements and the model of the desired signal. As the clipped sample reconstruction is generally converted to an estimation problem of exponential components, it can be implemented with root-MUltiple SIgnal Classification (root-MUSIC), Prony's method \cite{Hayes1996}, etc. To more efficiently and accurately reconstruct the cut-out samples, we suggest using matrix pencil method in this paper, which leads to the proposed matrix-pencil method based interference mitigation.

\subsection{Interference mitigation}
The flowchart of the matrix-pencil method based interference mitigation for FMCW radars is shown in Fig.~\ref{fig:Flowchart}. The detailed processing involves two main steps: 
\subsubsection{Interference detection and cutting out}
Based on the analysis in the previous section, the beat frequencies of targets are generally constant in a sweep while the interference after de-chirping and low-pass filtering still exhibits non-stationary spectral property within its duration. Taking advantage of this spectral difference, the interference and its duration can be detected with many approaches, such as energy spikers detection \cite{Kunert2012}, Constant False Alarm Rate (CFAR) thresholding \cite{Fischer2015}, complex baseband oversampling \cite{Murali2018} or other methods in time or time-frequency domain. After determining the location of the interference, the contaminated signal samples can be completely removed for interference suppression. However, it also eliminates part of the energy of the desired signals, which would cause signal to noise ratio (SNR) degradation of the resultant range profiles.   

\subsubsection{Signal extrapolation}
To overcome the SNR degradation of the targets' signals caused by the interference suppression, the removed signal samples can be reconstructed by using the interference-free samples and the corresponding signal model $\Tilde{s}_{\text{tar}}$. Generally, the all-pole signal model $\Tilde{s}_{\text{tar}}[k]$ is unknown and has to be estimated from the interference-free samples. In this paper, matrix-pencil method is applied to estimate the model parameters (i.e., model order, signal poles and the coefficients) by simultaneously accounting for the interference-free samples in front of and behind the clipped ones. Moreover, to alleviate the impact of the noise and signal discontinuity of the interference-free samples on the estimation of signal model of targets, an iterative fusion process is introduced to minimize the estimation error of the signals on the both sides of the clipped region relative to the interference-free measurements. If the estimation error fulfills a desired requirement after a few iterations, the signals in the cut-out region are reconstructed.

\begin{figure}[!t]
	\centering
	\includegraphics[width=0.43\textwidth]{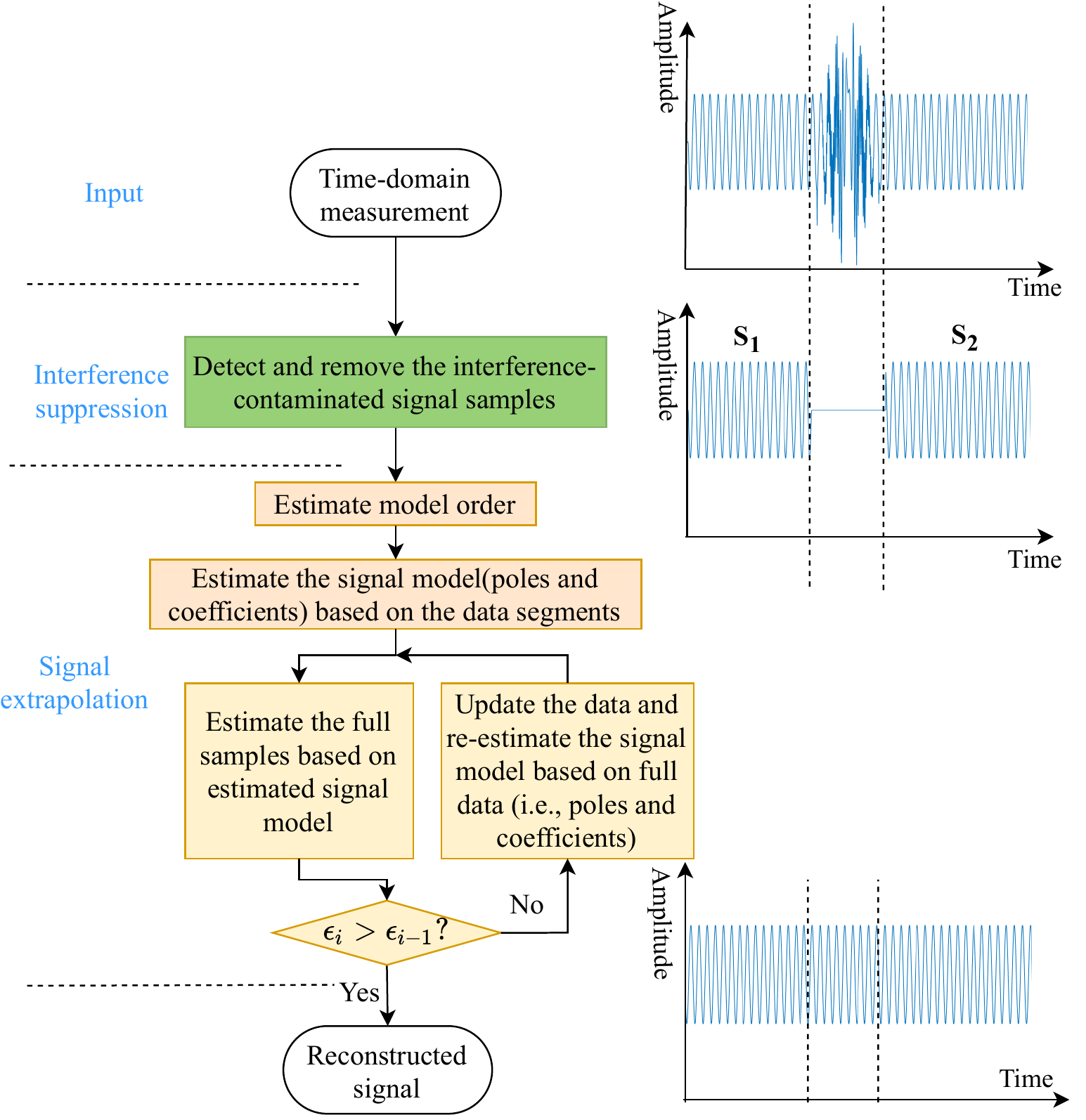}
	\caption{Flowchart of the proposed MP-based interference mitigation approach.}
	\label{fig:Flowchart}
\end{figure}

\subsection{Signal fusion and reconstruction}

After cutting out the interference-contaminated samples indexed from $N_1$ to $N_2$, the interference-free measurements in \eqref{Eq: discrete_signal_timeDom} can be represented as 
\begin{equation} \label{eq:signal_clipped}
    s[k] = \sum_{i=1}^{M^\prime} a_i z_i^k + n[k]
\end{equation}
where $k = 0,1,\cdots, N_1-1, N_2+1, N_2+2,\cdots, N-1$. Therefore, a gap is formed between the two signal sample segments from 0 to $N_1-1$ and from $N_2+1$ to $N$, as illustrated in the second plot on the right side of Fig.~\ref{fig:Flowchart}. As the useful signals in this gap are also eliminated due to the interference clipping, it would cause some SNR loss of the final coherent processing results (e.g., range profile, range-Doppler map, etc.). To overcome this problem, in the next step we try to reconstruct the useful signals in the gap based on the signal model \eqref{eq:signal_clipped} and the interference-free measurements on the both sides.   

As mentioned in the introduction, here the signal reconstruction can be converted to a signal fusion problem. We suggest using the matrix-pencil based fusion method in \cite{Zou2016,Wang2018} to implement the signal reconstruction but no incoherence correction between different signal segments is needed. 

For the convenience of description, we denote the signals before and after the clipped region as $s_1$ and $s_2$, given by 
\begin{equation}
    \left\{\begin{aligned}
    &s_1[k] = s[k], \quad k = 0, 1, \cdots, N_1-1\\
    &s_2[k] = s[k+N_2+1], \quad k = 0, 1, \cdots, N-N_2-2
    \end{aligned} \right.
\end{equation}
Then, the detailed steps of the signal reconstruction are presented as follows.

(1) Estimate the all-pole signal model \eqref{eq:signal_clipped} with the matrix pencil method based on the front and back signal segments, i.e., $s_1$ and $s_2$. 

Generally, the signal model order $M^\prime$ is estimated according to the Akaike Information Criterion (AIC), Bayesian Information Criterion (BIC), subspace-based automatic model order selection (SAMOS) \cite{Papy2007,Sun2018APMC}, etc. As SAMOS is considered to be one of the most general and robust approach to model order selection and outperforms the aforementioned methods based on the information theoretic criterion, it is used in this paper. The signal poles can be estimated with the matrix pencil method. Different from the signal pole estimation with continuous uniform signal samples, the Hankel matrices based on the discontinuous signals $s_1$ and $s_2$ are constructed in a slightly different way \cite{Wang2018,Zou2016}. Firstly, two Hankel matrices are constructed as
\begin{equation}
    \begin{aligned}
    & \mathbf{H}_{i0} = & [\mathbf{D}_{0}^i, \mathbf{D}_{1}^i, \cdots, \mathbf{D}_{L-1}^i]\\
    & \mathbf{H}_{i1} = & [\mathbf{D}_1^i, \mathbf{D}_2^i, \cdots, \mathbf{D}_L^i], 
    \end{aligned}
\end{equation}
with
\begin{equation}
    \mathbf{D}^i_k = [ s_i[k], s_i[k+1], \cdots, s_i[M_i-L-1+k] ]^T, \quad i = 1, 2.
\end{equation}
where $^T$ denotes the transpose operation, $M_1 = N_1$ and $M_2 = N-N_2-1$ are the lengths of $s_1$ and $s_2$, respectively. $L$ is the matrix pencil parameter and $\hat{M}^\prime<L< \min(M_1-\hat{M}^\prime, M_2 - \hat{M}^\prime)$, where $\hat{M}^\prime$ is the estimated signal model order (Without explicit statement, the $\hat{\cdot}$ notation represents the estimated value of a corresponding parameter).

The Hankel matrices constructed above can be vertically stacked as
\begin{equation}
    \mathbf{X}_{0} = 
    \left[\begin{array}{c} 
    \mathbf{H}_{10} \\ 
    \mathbf{H}_{20}
\end{array}\right], \quad
\mathbf{X}_{1} = 
    \left[\begin{array}{c} 
    \mathbf{H}_{11} \\ 
    \mathbf{H}_{21}
\end{array}\right].
\end{equation}
Then the matrix pencil $\mathbf{L}(\lambda) = \mathbf{X}_1-\lambda \mathbf{X}_0$ can be evaluated to get the estimates the signal poles $z_i$ in \eqref{eq:signal_clipped} \cite{Wang2018,Zou2016}. To get the eigenvalues of this matrix pencil, we take advantage of the singular value decomposition (SVD)-based method in \cite{Sarkar1995}. Taking the SVD of the matrix $\mathbf{X}_0$ and $\mathbf{X}_1$, we get
\begin{align} \label{eq:X0_SVD}
\mathbf{X}_0 = [\mathbf{U}_0, \mathbf{U}_0^\prime]\left[\begin{array}{cc}
\mathbf{\Sigma}_{0,\hat{M}^\prime} & 0 \\
0 & \mathbf{\Sigma}_{0,L-\hat{M}^\prime}
\end{array} \right] [\mathbf{V}_0, \mathbf{V}_0^\prime]^H \\   \label{eq:X1_SVD}
\mathbf{X}_1 = [\mathbf{U}_1, \mathbf{U}_1^\prime]\left[\begin{array}{cc}
\mathbf{\Sigma}_{1,\hat{M}^\prime} & 0 \\
0 & \mathbf{\Sigma}_{1,L-\hat{M}^\prime}
\end{array} \right] [\mathbf{V}_1, \mathbf{V}_1^\prime]^H 
\end{align}       
where $^H$ denotes the conjugate transpose of a matrix, $\mathbf{\Sigma}_{0,\hat{M}^\prime}$ and $\mathbf{\Sigma}_{1,\hat{M}^\prime}$ are the diagonal matrices containing $\hat{M}^\prime$ dominant singular values of $\mathbf{X}_0$ and $\mathbf{X}_1$, respectively. The columns of $\mathbf{U}_0$, $\mathbf{U}_1$, $\mathbf{V}_0$ and $\mathbf{V}_1$ are the left and right singular vectors related to the dominant singular values. $(\mathbf{U}_0, \mathbf{\Sigma}_{0,\hat{M}^\prime}, \mathbf{V}_0 )$ and $(\mathbf{U}_1, \mathbf{\Sigma}_{1,\hat{M}^\prime}, \mathbf{V}_1 )$ are the singular value systems related to the signal subspace in $\mathbf{X}_0$ and $\mathbf{X}_1$,respectively. The rest terms in \eqref{eq:X0_SVD} and \eqref{eq:X1_SVD} form the corresponding singular value systems related to the so-called noise subspace.

To suppress the impact of the noise on the signal pole estimation, $\mathbf{X}_0$ and $\mathbf{X}_1$ can be approximated by their truncated SVD as $\mathbf{X}_{0T}$ and $\mathbf{X}_{1T}$
\begin{align}
\mathbf{X}_0 \approx \mathbf{X}_{0T} = \mathbf{U}_0 \mathbf{\Sigma}_{0,\hat{M}^\prime} \mathbf{V}_0^H \\
\mathbf{X}_1 \approx \mathbf{X}_{1T} = \mathbf{U}_1 \mathbf{\Sigma}_{1,\hat{M}^\prime} \mathbf{V}_1^H
\end{align}
Then the signal poles $z_i$ can be estimated by solving the generalized eigenvalue problem $\det\left(\mathbf{L}(\lambda)\right)=0$ of the matrix pair $\{\mathbf{X}_0; \mathbf{X}_1\}$, which is equivalent to the ordinary eigenvalue problem 
\begin{equation} 
\det \left( \mathbf{\Sigma}_{0,\hat{M}^\prime}^{-1} \mathbf{U}_0^H \mathbf{U}_1 \mathbf{\Sigma}_{1,\hat{M}^\prime} \mathbf{V}_1^H \mathbf{V}_0 - \lambda  \mathbf{I}  \right)=0
\end{equation}
The signal pole estimations $\hat{z}_i = \lambda_i,\,i=1,2,\cdots, \hat{M}^\prime$ are obtained. 

After that, using the estimated signal model order $\hat{M}^\prime$ and the signal poles $\hat{z}_i$, the complex amplitude $a_i$ can be cast as the least-square problem $\mathbf{m} =\mathbf{Z}\mathbf{a}$, where $\mathbf{m} = [s_1, s_2]^T$ is the measured interference-free data, $\mathbf{Z}$ is the matrix formed by signal poles and $\mathbf{a}=[a_1, a_2, \cdots, a_{\hat{M}^\prime}]$ is the vector of the coefficients. Explicitly, it is represented as
\begin{small}
\begin{equation}
\left[\begin{array}{c}
s_1[0] \\ s_1[1] \\ \vdots \\ s_1[M_1-1] \\ s_2[0] \\  \vdots \\ s_2[M_2-1]
\end{array}
\right]= \left[ \begin{array}{cccc}
1 & 1 & \cdots & 1 \\
z_1 & z_2 & \cdots & z_{\hat{M}^\prime} \\
\vdots & \vdots & \ddots & \vdots \\
z_1^{N_1-1} & z_2^{N_1-1} & \cdots & z_{\hat{M}^\prime}^{N_1-1} \\
z_1^{N_2} & z_2{N_2} & \cdots & z_{\hat{M}^\prime}^{N_2} \\
\vdots & \vdots &\ddots & \vdots \\
z_1^{N-1} & z_1^{N-1} & \cdots & z_1{N-1}
\end{array} 
\right]\left[ \begin{array}{c}
a_1 \\ a_2 \\ \vdots \\ a_{\hat{M}^\prime}
\end{array}\right]
\end{equation} 
\end{small}

(2) After inserting the estimated signal poles $\hat{z}_i$ and the coefficients $\hat{a}_i$ into \eqref{eq:signal_clipped}, the full beat signal in the sweep can be estimated by
\begin{equation} \label{eq:sigMod_fullBeatSig}
\hat{s}[k] = \sum_{i=1}^{\hat{M}^\prime} \hat{a}_i \hat{z}_i^k, \qquad k=0, 1, \cdots, N-1
\end{equation}
The estimated full beat signal indicates 
\begin{equation}
\left\{ \begin{aligned}
\hat{s}_1[k] &= \hat{s}[k], \qquad k\in [0, N_1-1] \\
\hat{s}_g[k-N_1] &= \hat{s}[k], \qquad k \in [N_1, N_2] \\
\hat{s}_2[k-N_2-1] &= \hat{s}[k], \qquad k \in [N_2+1, N-1]
\end{aligned}  \right.
\end{equation}

(3) To improve the estimation of the full beat signal, we replace the $\hat{s}_1$ and $\hat{s}_2$ parts in $\hat{s}$ with the measurements $s_1$ and $s_2$. Then the reconstructed full beat signal can be modified as
\begin{equation} \label{eq:sig_reconstr_mix_est_measure}
\hat{s}[k] = \left\{ \begin{aligned}
&s_1[k],  &\qquad  k \in [0, N_1-1] \\
&\hat{s}_g[k-N_1], &\qquad  k \in [N_1, N_2] \\
&s_2[k-N_2-1], &\qquad k \in [N_2+1, N-1]
\end{aligned} \right.
\end{equation}

Next, the reconstructed signal $\hat{s}$ in \eqref{eq:sig_reconstr_mix_est_measure} are used as a set of contiguous samples to re-estimate the signal poles $z_i$ and the coefficients $a_i$ in \eqref{eq:signal_clipped} by using the traditional matrix-pencil method \cite{Sarkar1995}.    
 
(4) Repeat steps (2) and (3) to update the reconstructed results. After the step (2) in each iteration, the $l^2$-norm of the differences between the estimated signals and their measured counterparts is examined to quantify the signal estimation accuracy
\begin{equation}
\epsilon_i = \| \hat{s}^{(i)}_1 - s_1  \|_2 + \| \hat{s}^{(i)}_2 - s_2 \|_2,
\end{equation}
where $\hat{s}^{(i)}_1$ and $\hat{s}^{(i)}_2$ are the estimated counterparts of the measurements $s_1$ and $s_2$ in the $i^\text{th}$ iteration. If the signal difference in the $i^\text{th}$ iteration satisfies the requirement 
\begin{equation}
\epsilon_i > \epsilon_{i-1},
\end{equation}
then iteration will stop. Otherwise, it continues to improve the estimated model parameters. 

After several iteration cycles, we get the most accurate recovery of the full beat signal. Finally, by taking corresponding operations on the reconstructed full beat signal, the range profile and Doppler information of targets can be obtained with substantially improved dynamic range and suppressed ``noise'' floor.

\section{Numerical Simulations} \label{sec: Num_Simu}

{ \renewcommand{\arraystretch}{1.1}
\begin{table}[!t]
    \centering
    \caption{Parameters used for simulations for point-like and distributed target scenarios}
    \label{Table: Simulation Parameters point-like and Distibuted}
\begin{tabular}{|l|l|l|}
\hline
\textbf{Parameter}          & \textbf{Value}  & \textbf{Unit} \\ \hline
Center frequency                    & 3                               & GHz                            \\ \hline
Bandwidth                           & 40                              & MHz                            \\ \hline
FMCW sweep duration                          & 500                             & $\mu s$                             \\ \hline
Sweep slope                         &  $8\times 10^{10}$  & Hz/s    \\ \hline
Transmit Power                      & 1                               & Watt                           \\ \hline
Sampling frequency                  & 12                              & MHz                            \\ \hline
Maximum unambiguous range                       & 8                           & km                              \\ \hline
\multicolumn{3}{|c|}{\textbf{Point target scenario}}                            \\ \hline
Distances of three targets                    & 2, 5, and 5.1                            & km                              \\ \hline
Interference duration & 10-50\% & N/A \\
\hline 
\multicolumn{3}{|c|}{\textbf{Extended targets scenario}}                           \\ \hline
Number of point targets                   & 15                              & N/A                            \\ \hline
Distance between adjacent targets   & $1<d<1.8$                       & m                              \\ \hline
\multirow{2}{4.5cm}{Interference duration relative  to the sweep duration} & \multirow{2}{*}{$24.3\%$} & \multirow{2}{*}{N/A} \\ 
& & \\
\hline 
\end{tabular}
\end{table}
}

To analyze performance of the proposed MP-based method to interference mitigation, several sensing scenarios have been simulated. Its results are also compared with the traditional zeroing and two of the state-of-the-art methods, i.e., Burg-based approach \cite{Neemat2019} and the IVM-based method \cite{Toth2019RadarConf}.

\subsection{Evaluation metric}
To facilitate quantitative evaluation of the accuracy of the reconstructed beat signals by different methods, we introduce two evaluation metrics: the Relative Signal-to-Noise Ratio (RSNR) and the correlation coefficient $\rho$. The RSNR and the correlation coefficient are defined as
\begin{align} \label{eq:relative_SNR}
\text{RSNR}(\mathbf{s}_0, \hat{\mathbf{s}}) &= 20\log_{10}\frac{ \left\|\mathbf{s}_0\right\|_2}{\left \|\mathbf{s}_0-\hat{\mathbf{s}}\right\|_2} \\ 
\label{eq:corr_coefficient}
\rho_{\mathbf{s}_0, \hat{\mathbf{s}}} &= \frac{\hat{\mathbf{s}}^H \mathbf{s}_0}{\left \|\mathbf{s}_0\right\|_2 \cdot \left \|\hat{\mathbf{s}}\right\|_2}
\end{align}
where $\mathbf{s}_0$ is the vector of a clean reference beat signal (without interferences and noise) and  $\hat{\mathbf{s}}$ is the beat signal formed by the measured interference-free samples and the reconstructed signal samples in the cut-out region. $\left\|\cdot\right\|_2$ denotes the $\ell^2$ norm operator. If the signal samples in the cut-out region are reconstructed with sufficient accuracy, a RSNR larger than the SNR of the input signal can be obtained according to \eqref{eq:relative_SNR}. So the larger the obtained RSNR is, the more accurate the recovered signal samples are.

The correlation coefficient is commonly used to evaluate the similarity of two signals. Its formulation in \eqref{eq:corr_coefficient} is a normalized inner product between the reconstructed signal and the reference one, which specifically represents the rotation angle between the two signals. The correlation coefficient satisfies $0 \leq |\rho_{\mathbf{s}_0, \hat{\mathbf{s}}}| \leq 1$. If $ |\rho_{\mathbf{s}_0, \hat{\mathbf{s}} }| = 1$, then the reconstructed signal $\hat{\mathbf{s}}$ is a linear function of the reference signal $\mathbf{s}_0$ with phase difference of $\angle \rho_{\mathbf{s}_0, \hat{\mathbf{s}}}$ (i.e., argument of $\rho_{\mathbf{s}_0, \hat{\mathbf{s}}}$). That is to say,  a correlation coefficient with a larger modulus and a smaller argument indicates a better recovery performance.

\subsection{Point target scenario} \label{subsec:point_target_scenario}
Firstly, we demonstrate the performance of the proposed MP-based interference mitigation approach in the point target scenario. The parameters of the FMCW radar system used for the simulation are shown in Table~\ref{Table: Simulation Parameters point-like and Distibuted}. Three point targets are placed at a distance of $2\,\mathrm{km}$, $5\,\mathrm{km}$ and $5.1\,\mathrm{km}$, respectively, away from the transceiver. The amplitudes of the scattered signals from the three targets from the near to further distances are set to be 1, 0.2, and 0.1, respectively.  

\begin{figure}[t!]
    \centering
    \subfloat[]{
    \includegraphics[width=0.45\textwidth]{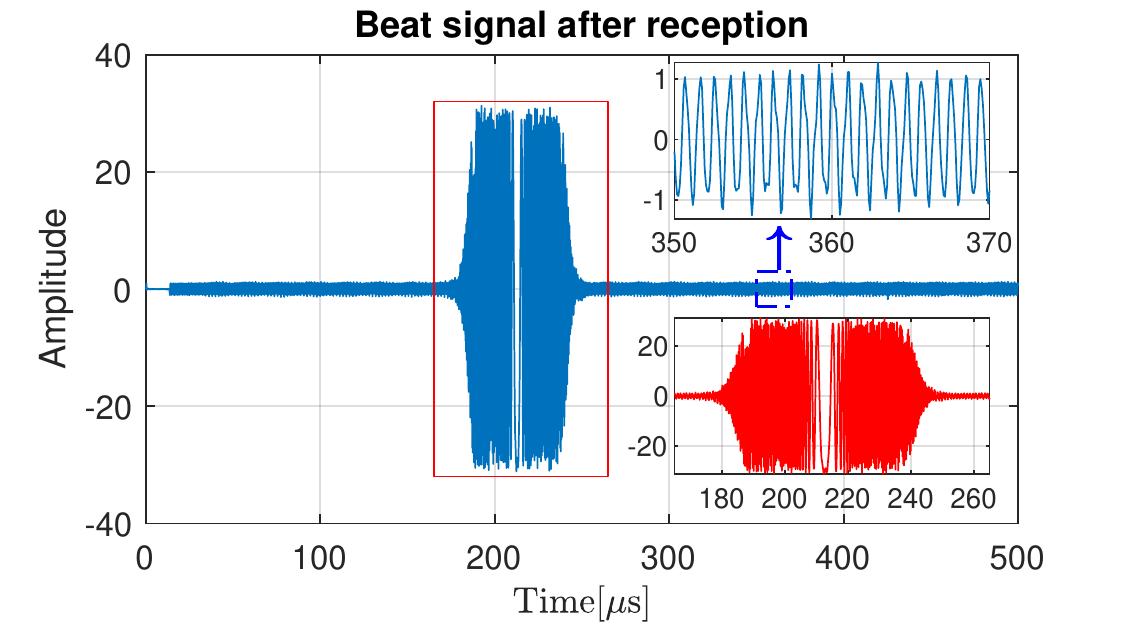}
    \label{fig:PtTar_SNR15_BeatSignal}
    }

    \vspace{-3mm}
    \subfloat[]{
    \includegraphics[width=0.4\textwidth]{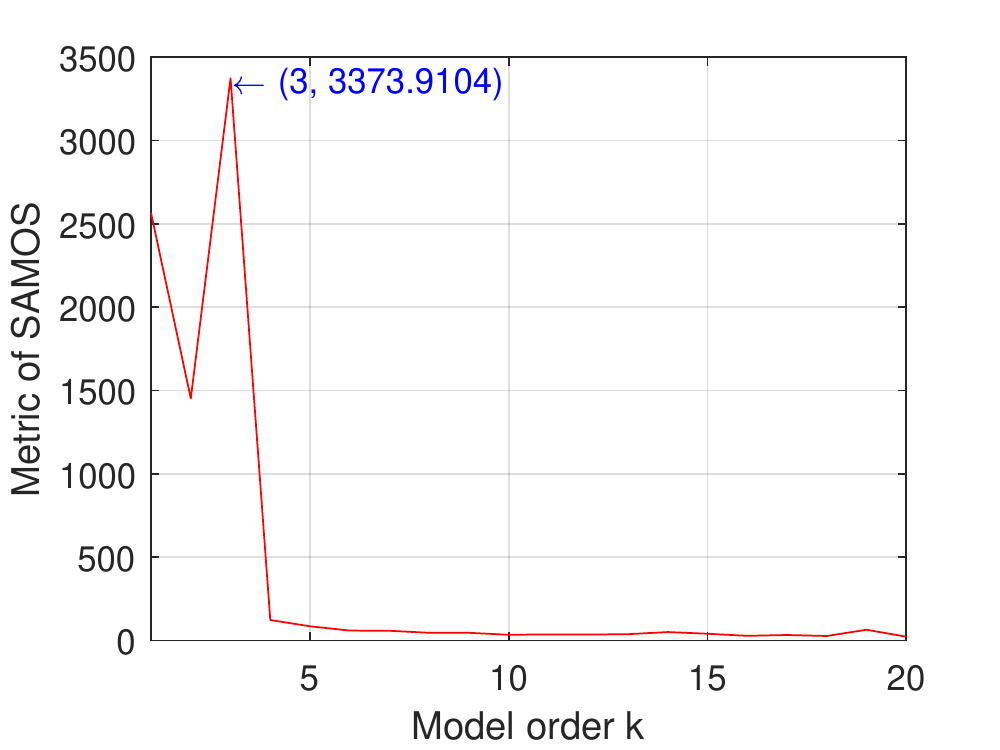}
    \label{fig:PtTar_SNR15_modOrder}
    }
    
    \vspace{-2mm}
    \centering
    \subfloat[]{
    \includegraphics[width=0.45\textwidth]{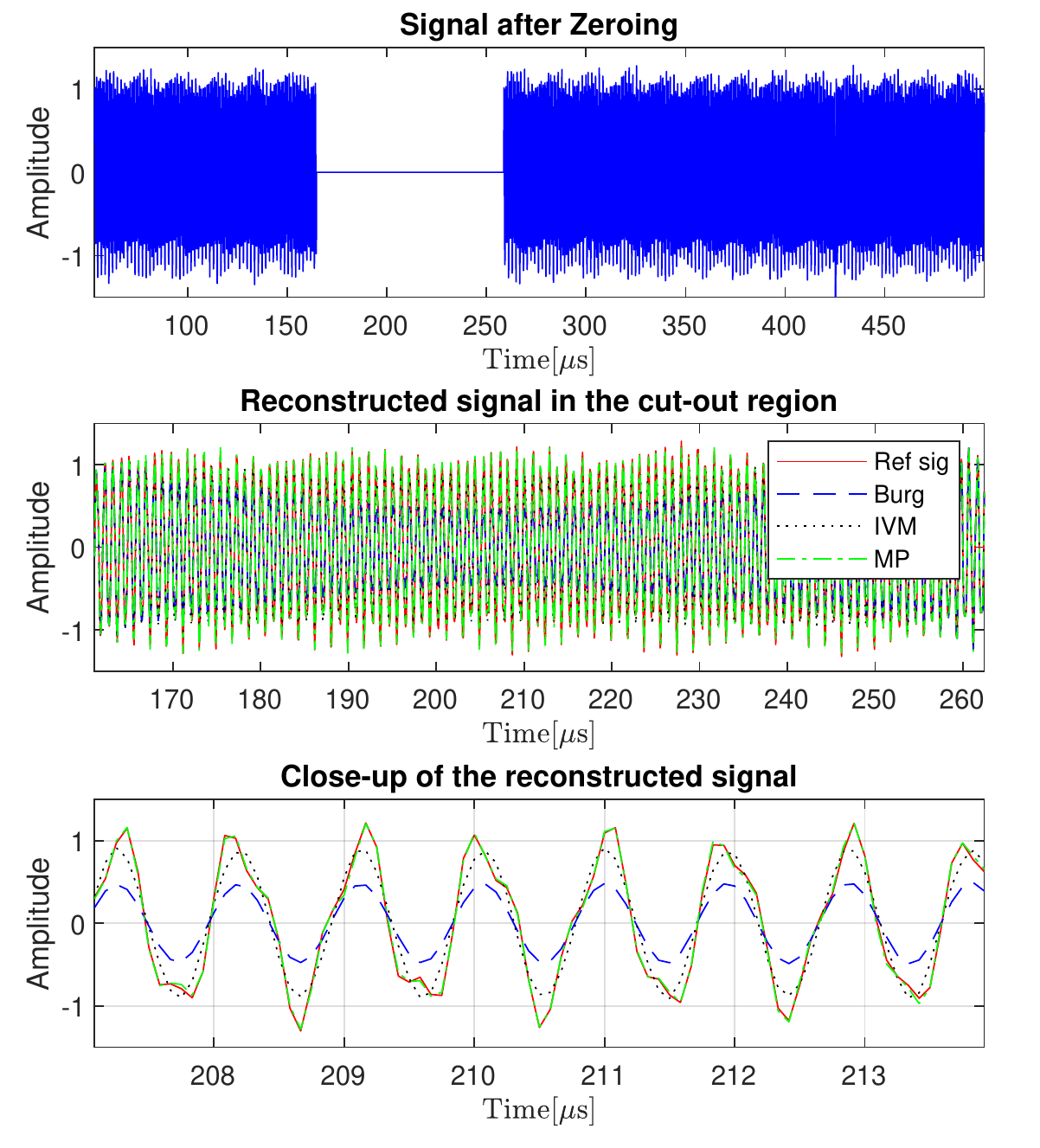}
    \label{fig:PtTar_SNR15_RawSig}
    }

    \caption{ Numerical simulation for interference mitigation in the point target scenario. \protect\subref{fig:PtTar_SNR15_BeatSignal} shows the interference-contaminated beat signal. \protect\subref{fig:PtTar_SNR15_modOrder} presents the metric values of SAMOS for model order estimation.  \protect\subref{fig:PtTar_SNR15_RawSig} displays the results after interference mitigation.  }
    \label{fig:Simu_PtTar_SNR15}
\end{figure}

The victim FMCW radar system suffers from a strong interference from an aggressor FMCW radar with the same operational center frequency but an opposite sweep slope and a time advancement of $75\mu s$ relative to the starting time of the victim sweep. After dechirping, the interference-contaminated beat signal is acquired and illustrated in Fig.~\ref{fig:Simu_PtTar_SNR15}\subref{fig:PtTar_SNR15_BeatSignal}. The strong interference appears at the interval from $165\,\mathrm{\mu s}$ to $265\,\mathrm{\mu s}$ (indicated by the red solid-line rectangle), which still exhibits as a chirp-like signal (see the bottom-right inset in Fig.~\ref{fig:Simu_PtTar_SNR15}\subref{fig:PtTar_SNR15_BeatSignal}). Meanwhile, for clarity, part of the interference-free beat signal (from $350\,\mathrm{\mu s}$ to $370\,\mathrm{\mu s}$ indicated by the blue dash-dotted rectangle) is zoomed in and shown in the top-right inset. It is clear that the beat signal of targets is composed of the sinusoidal components. Moreover, white Gaussian noise with the SNR of $15\,\mathrm{dB}$ is added to the signal to account for the thermal noise and measurement errors of the radar system.

\begin{figure}
    \centering
    \vspace{-2mm}
    \subfloat[]{
    \includegraphics[width=0.425\textwidth]{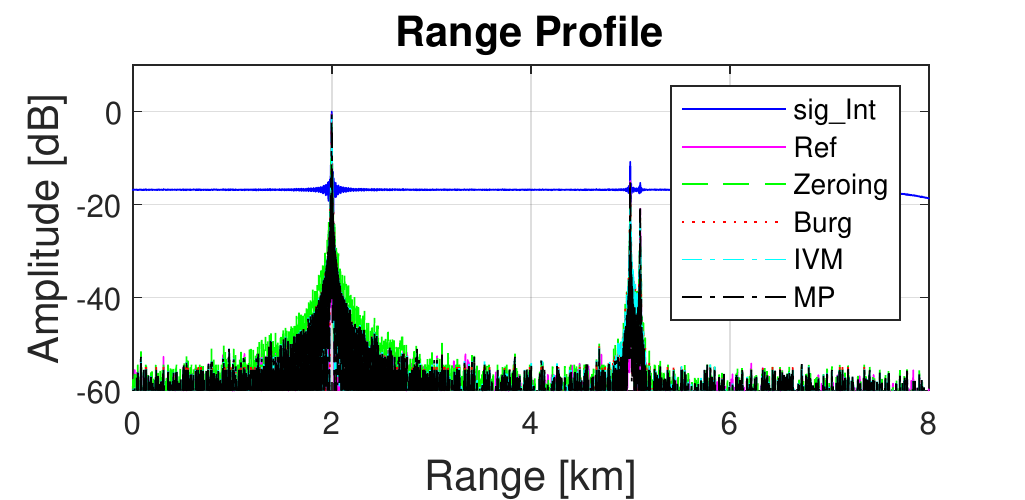}
    \label{fig:PtTar_SNR15_RangeProfile}
    }
    
    \vspace{-3mm}
    \subfloat[]{
    \includegraphics[width=0.24\textwidth]{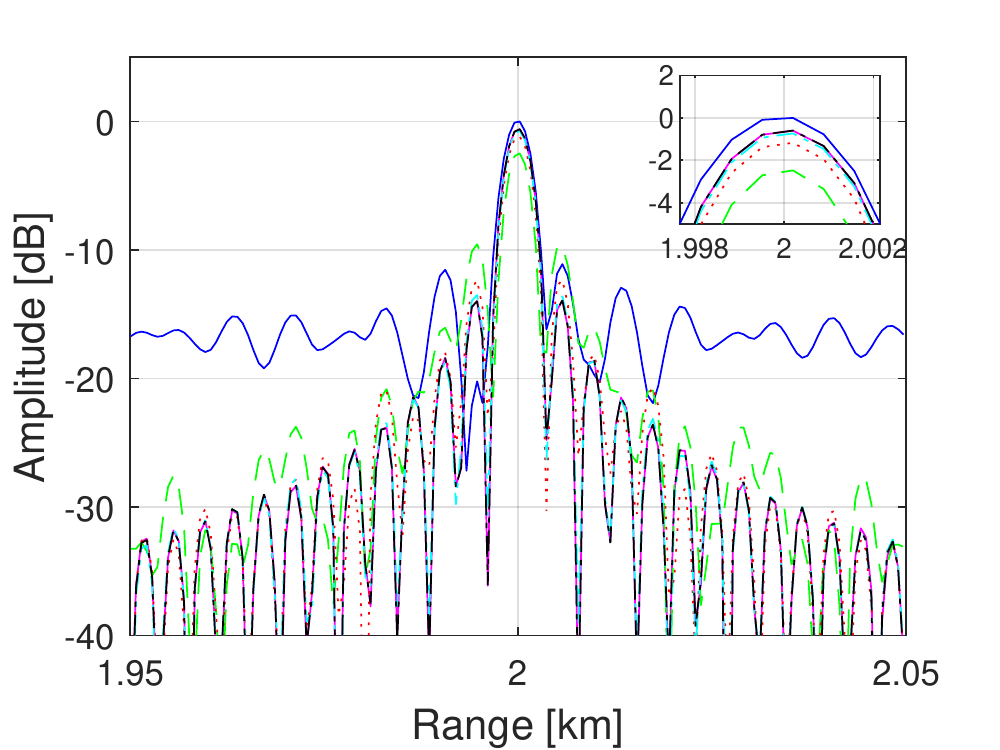}
    \label{fig:PtTar_SNR15_RProfile_close-up_2km}
    }
    \subfloat[]{
    \includegraphics[width=0.24\textwidth]{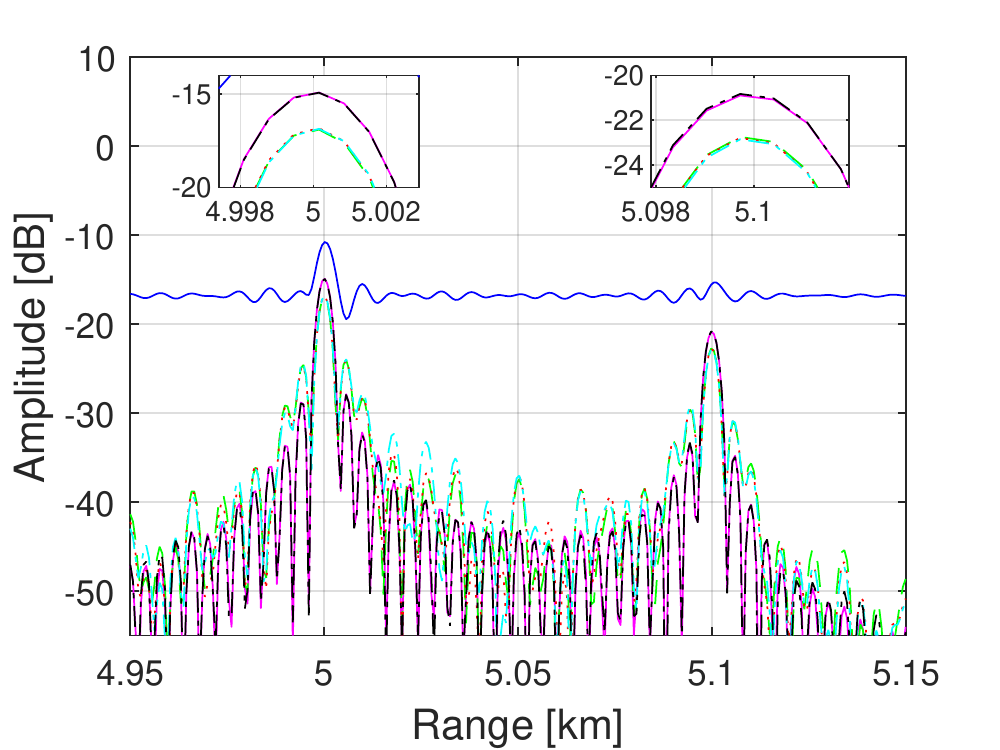}
    \label{fig:PtTar_SNR15_RProfile_close-up_5km}
    }
    \caption{\protect\subref{fig:PtTar_SNR15_RangeProfile} displays the range profiles of the targets obtained with the interference-contaminated signal, interference-free reference signal, and the signals processed with the zeroing, Burg- and MP-based methods; \protect\subref{fig:PtTar_SNR15_RProfile_close-up_2km} and \protect\subref{fig:PtTar_SNR15_RProfile_close-up_5km} are the close-ups of the range profile around the distances of targets, respectively.}
    \label{fig:Simu_PtTar_SNR15_RangeProfile}
\end{figure}

The interference-contaminated beat signal produces a range profile with significantly increased noise floor (see ``sig\_Int'' in Fig.~\ref{fig:Simu_PtTar_SNR15_RangeProfile}\subref{fig:PtTar_SNR15_RangeProfile} where the two targets at the further distances are almost shadowed by the raised noise floor) if the range compression is performed directly by using the fast Fourier transform (FFT). To mitigate the interference by using the proposed MP-based approach, the interference-contaminated samples of the signal are firstly detected and cut out (i.e., zeroing with a rectangular window \cite{Babur2010}). Zeroing the interference-contaminated samples results in two separate signal segments with a gap inbetween (see the top panel in Fig.~\ref{fig:Simu_PtTar_SNR15}\subref{fig:PtTar_SNR15_RawSig}), which causes not only power loss of targets' signals but also high sidelobes of the range profile, thus degrading the performance of target detection. To overcome these effects, the proposed MP-based interference mitigation method is used to reconstruct the signal samples in the cut-out gap based on the signal model \eqref{eq:sigMod_fullBeatSig} and the rest interference-free ones in front and back. Before reconstruction, the model order was estimated to be three by using the SAMOS method (see Fig.~\ref{fig:Simu_PtTar_SNR15}\subref{fig:PtTar_SNR15_modOrder}), which agrees with the true value. Then, by exploiting the proposed iterative scheme, the signal samples in the gap were recovered with sufficient accuracy, as shown in the middle plot and a close-up of them in the bottom panel in Fig.~\ref{fig:Simu_PtTar_SNR15}\subref{fig:PtTar_SNR15_RawSig}. For comparison, the interference-free reference signal (with the noise) and the recovered signals with the Burg-based method and the IVM, which used the same model order as that of the MP-based method, are also shown in the middle and bottom panels. One can see that both the signal recovered with the proposed MP-based method has the best agreement with the reference one. Meanwhile, the IVM method achieves more accurate reconstruction of the signals in the cut-out region than the Burg-based method in this case.             

To further examine the accuracy of the reconstructed signals, the range profiles of targets are constructed by taking the FFT of them and shown in Fig.~\ref{fig:Simu_PtTar_SNR15_RangeProfile}. For comparison, the range profiles obtained with the interference-contaminated [i.e., ``sig\_Int'' in Fig.~\ref{fig:Simu_PtTar_SNR15_RangeProfile}\subref{fig:PtTar_SNR15_RangeProfile}] and interference-free reference beat signals [i.e., ``Ref'' in Fig.~\ref{fig:Simu_PtTar_SNR15_RangeProfile}\subref{fig:PtTar_SNR15_RangeProfile}] are also presented. Note all the range profiles in Fig.~\ref{fig:Simu_PtTar_SNR15_RangeProfile} are normalized by the maximum of the range profile acquired with the interference-contaminated signal.  

According to Fig.~\ref{fig:Simu_PtTar_SNR15_RangeProfile}\subref{fig:PtTar_SNR15_RangeProfile}, all the interference mitigation methods, i.e., zeroing, Burg-, IVM- and MP-based methods, significantly reduce the ``noise'' floor of the range profile and thus increase its dynamic range compared to the one obtained with the interference-contaminated signal. Among them, the zeroing method is computationally most efficient by simply replacing the interference-contaminated samples with zeros, however, resulting in a gap between the front and rear signal samples. Consequently, it causes high side-lobes and some SNR loss in the range profile compared to that obtained with the reference signal. Specifically, from the insets in Fig.~\ref{fig:Simu_PtTar_SNR15_RangeProfile}\subref{fig:PtTar_SNR15_RProfile_close-up_2km} and \subref{fig:PtTar_SNR15_RProfile_close-up_5km}, the peaks of targets' range profiles obtained after zeroing are $1.9\,\mathrm{dB}$ lower than those formed with the reference signal and the signal reconstructed with the MP-based method. Although the Burg- and IVM-based method efficiently interpolate the samples in the cut-out gap and result in comparable/identical range profiles as the reference signal for the target at the short distance, they fail to overcome the power loss for the two weak targets at the further distances and get range profiles close to that of the zeroing method (see the insets in Fig.~\ref{fig:Simu_PtTar_SNR15_RangeProfile}\subref{fig:PtTar_SNR15_RProfile_close-up_5km}). By contrast, the MP-based method not only conquers the power loss of the range profile for all the targets but also accurately reconstructs their range profiles in terms of both main lobe and the side-lobes. Quantitatively, for the beat signals in Fig.~\ref{fig:Simu_PtTar_SNR15}\subref{fig:PtTar_SNR15_RawSig} recovered with the Burg-, IVM- and MP-based methods, their RSNRs are $14.55\,\mathrm{dB}$, $18.86\,\mathrm{dB}$ and $28.68\,\mathrm{dB}$, and the corresponding correlation coefficients are $0.9830e^{-j0.0018}$, $0.9935e^{j0.0044}$ and $0.9993e^{j0.0003}$, respectively, relative to the clean reference signal. Therefore, it recovers the signal samples in the cut-out region more accurately than the Burg- and IVM-based methods.

\subsection{Extended target scenario}

\begin{figure*}[!t]
	\centering
	\begin{minipage}{\textwidth}
	\centering
	\subfloat[]{\hspace{-4mm}
		\includegraphics[width=0.42\textwidth]{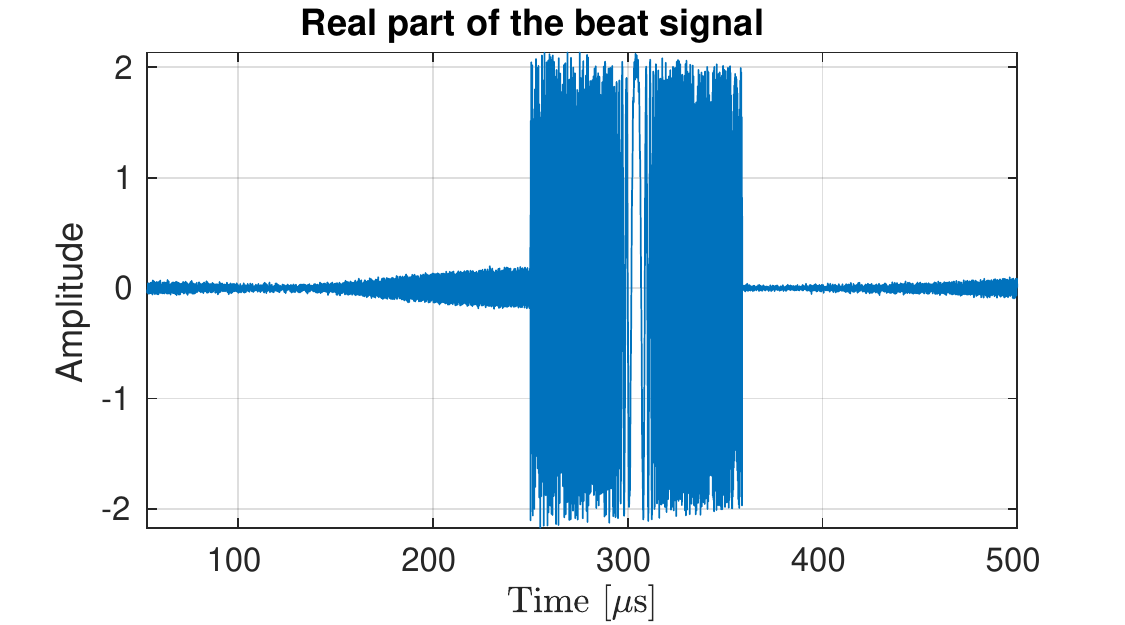}
		\label{fig:disTar_SNR15BeatSig_real}
	}
	\subfloat[]{ \hspace{-8mm}
	\includegraphics[width=0.31\textwidth]{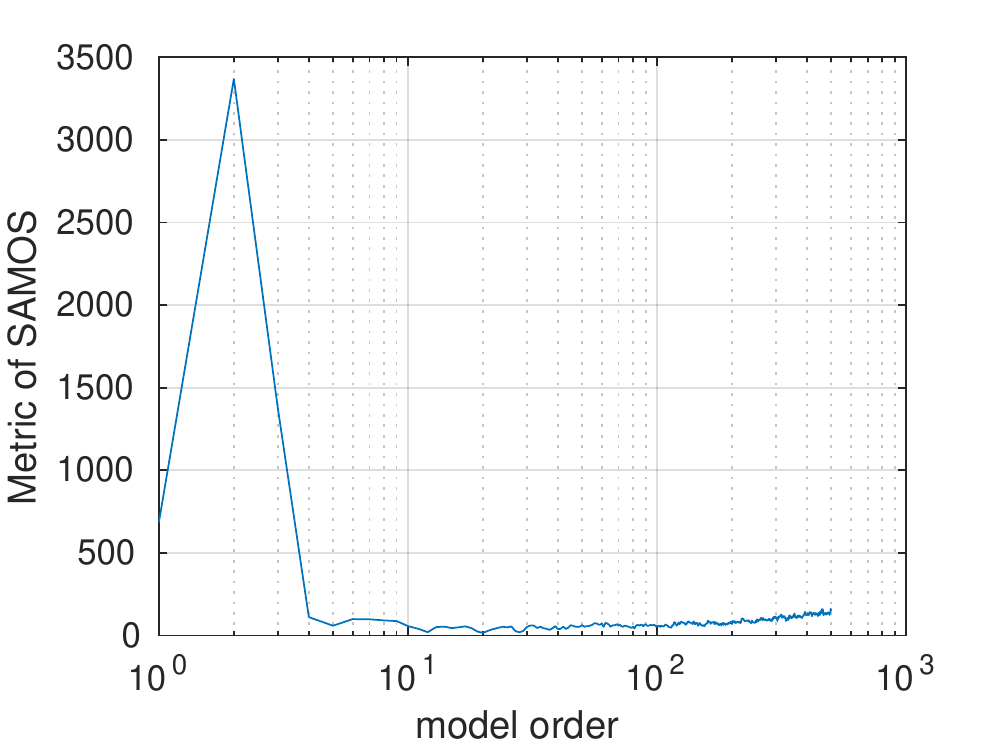}
	\label{fig:disTar_SNR15_mOd}
	}
	\subfloat[]{\hspace{-5mm}
	\includegraphics[width=0.31\textwidth]{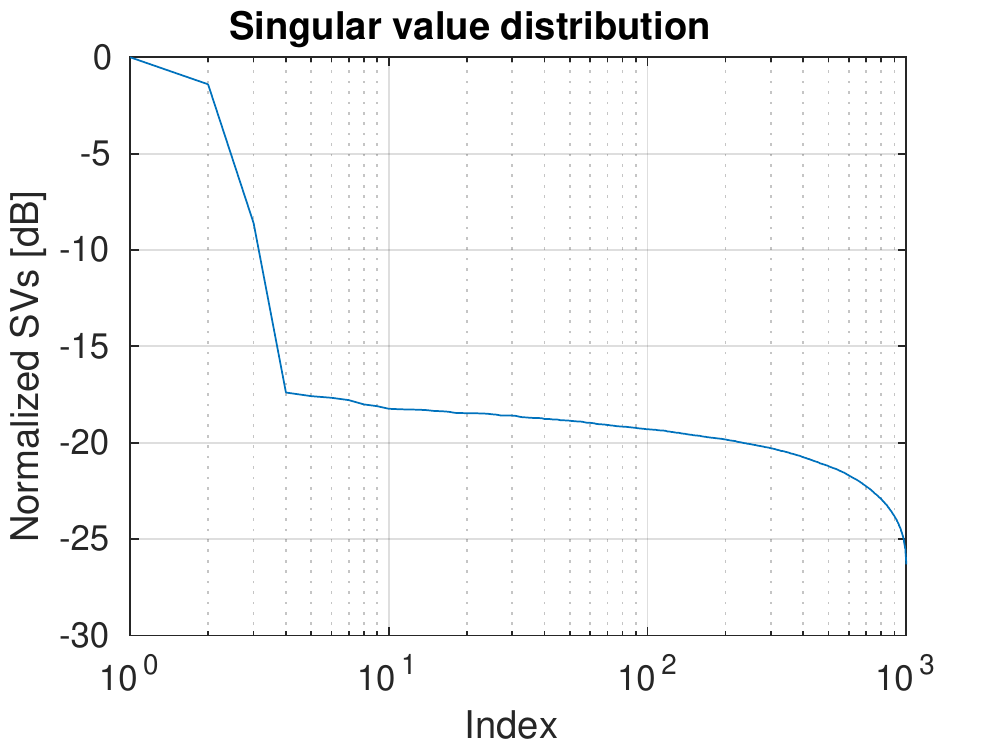}
	\label{fig:disTar_SNR_SV}
	}	
	\end{minipage}

	\begin{minipage}{\textwidth}
	\centering
	\begin{minipage}{0.5\textwidth}
	\centering
	\subfloat[]{
	\includegraphics[width=0.9\textwidth]{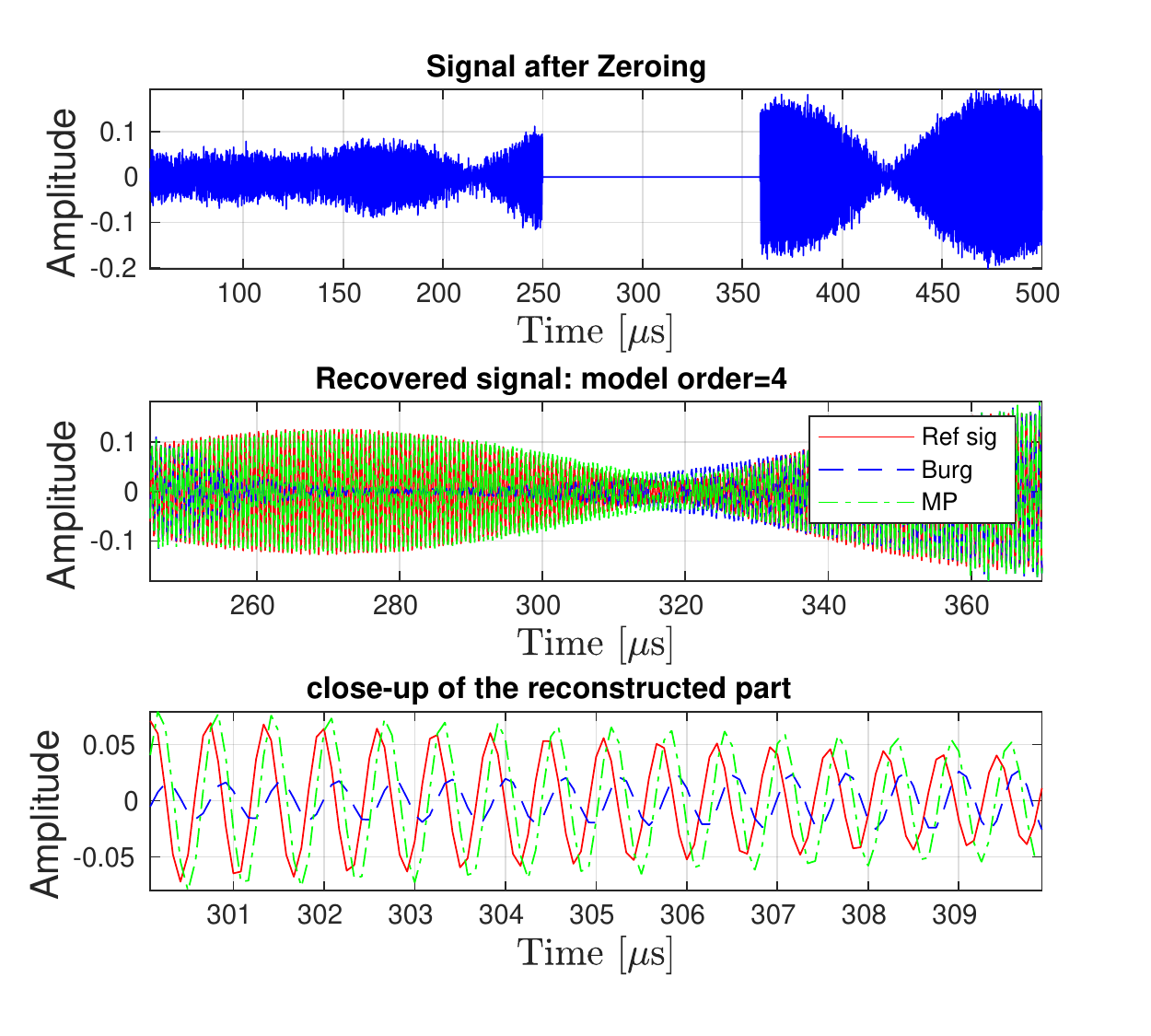}
	\label{fig:disTar_SNR15_RawSig_real_mOd4}
	}	
	\end{minipage}
    \begin{minipage}{0.48\textwidth}
    \centering
	\subfloat[]{
	\includegraphics[width=0.88\textwidth]{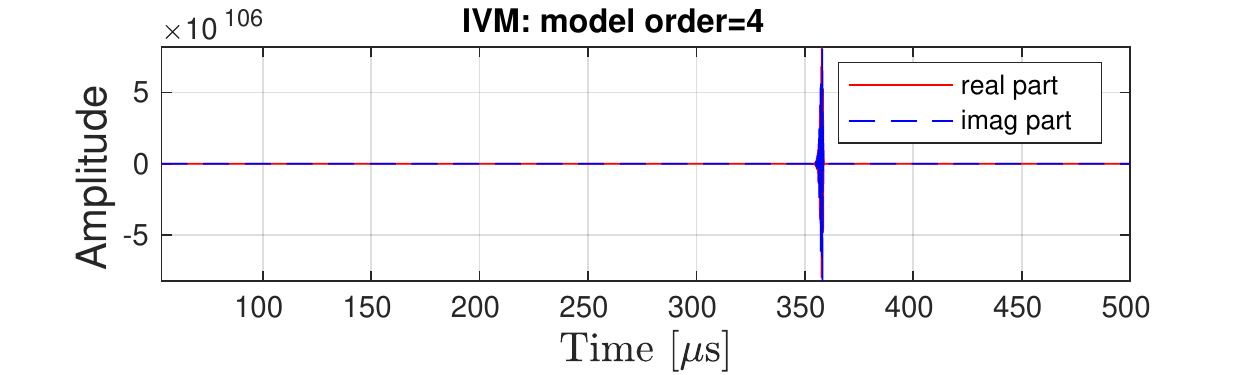}
	\label{fig:disTar_SNR15_RawSig_mOd4_IVM}  	
	}    
	
	\vspace{-3mm}    
	\subfloat[]{
		\includegraphics[width=0.88\textwidth]{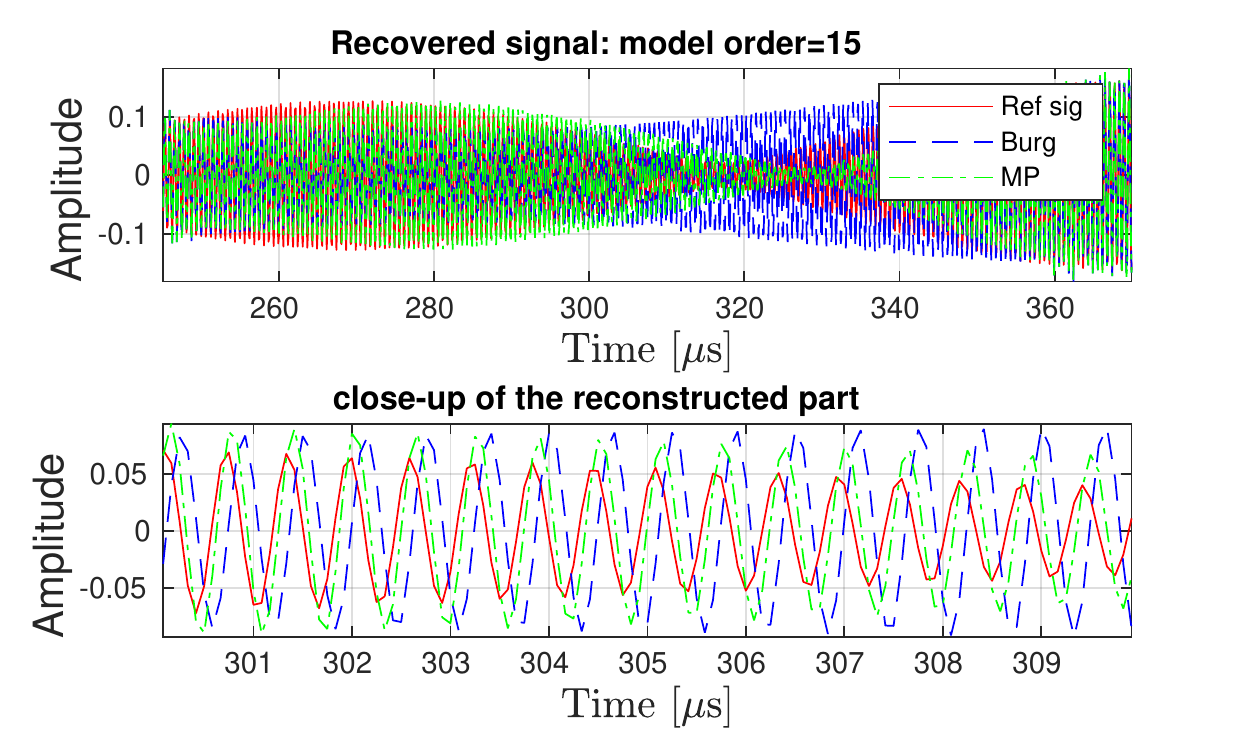}
		\label{fig:disTar_SNR15_RawSig_real_mOd15}	
	}
    \end{minipage}
	\end{minipage}	

	
\caption{Numerical simulation for interference mitigation in the extended target scenario. \protect\subref{fig:disTar_SNR15BeatSig_real} shows the interference-contaminated beat signal of an extended target. \protect\subref{fig:disTar_SNR15_mOd} shows the metric values of SAMOS approach for model order selection while \protect\subref{fig:disTar_SNR_SV} presents the singular values distribution of the matrix constructed for model order selection.  \protect\subref{fig:disTar_SNR15_RawSig_real_mOd4} shows the results after interference mitigation, where the top panel gives the beat signal after zeroing; the middle panel presents the reference beat signal and the beat signals recovered with the Burg- and MP-based methods with the model order of four; and the bottom panel is the close-up view of the recovered samples in the cut-out region. \protect\subref{fig:disTar_SNR15_RawSig_mOd4_IVM} shows the beat signal recovered by the IVM with the model order of four and \protect\subref{fig:disTar_SNR15_RawSig_real_mOd15} displays the recovered beat signals with the model order of 15. }
\label{fig:disTar_SNR15_Sig_TD}
\end{figure*}

\begin{figure}[!t]
	\centering
	\vspace{-3mm}
	\subfloat[]{
	\includegraphics[width=0.34\textwidth]{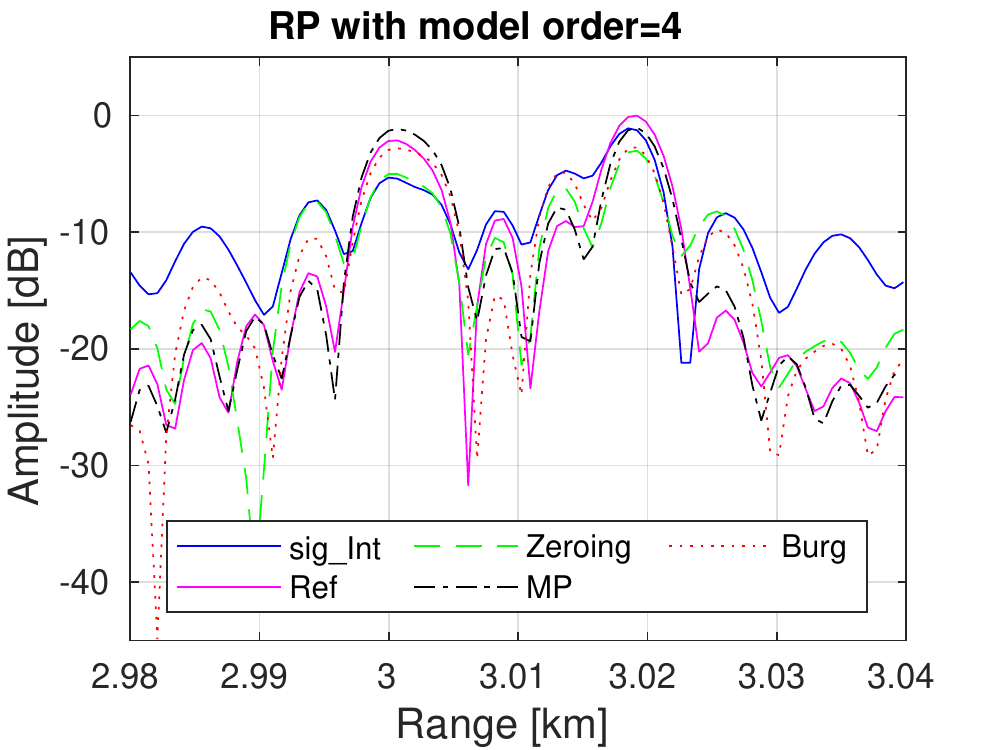}	
	\label{fig:disTar_SNR15_MP_RP_zoom_3km_mOd4}
	}

	\vspace{-3mm}
	\subfloat[]{
	\includegraphics[width=0.34\textwidth]{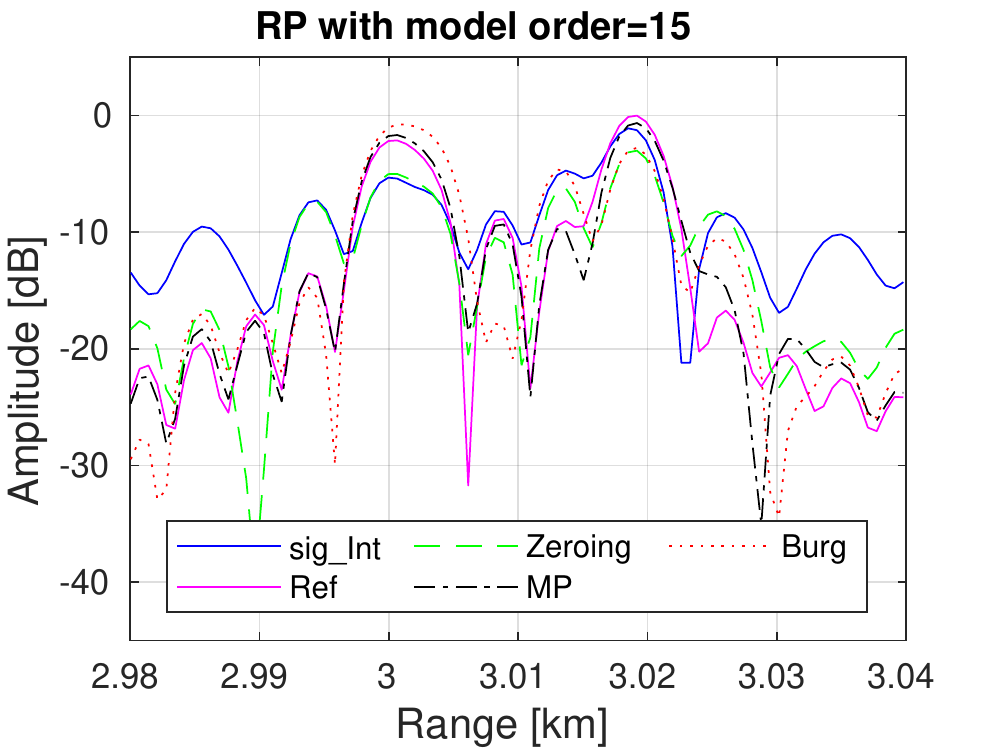}
	\label{fig:disTar_SNR15_MP_RP_zoom_3km_mOd15}
	}
	\caption{Range profile of the extended target obtained with the reference signal, interference-contaminated beat signal, the signals obtained with zeroing, and the signals recovered by the Burg- and MP-based methods with \protect\subref{fig:disTar_SNR15_MP_RP_zoom_3km_mOd4} the model order of four and \protect\subref{fig:disTar_SNR15_MP_RP_zoom_3km_mOd15} the model order of fifteen.}
	\label{fig:disTar_SNR15_MP_Range_zoom_3km}
\end{figure}

Here we consider the applicability of the proposed method to extended target scenarios. The parameters used for the simulation are shown in Table~\ref{Table: Simulation Parameters point-like and Distibuted}. An extended target formed by 15 point scatterers with adjacent inter-distances less than the range resolution of the radar system (i.e., $3.75\,\mathrm{m}$ in ours simulation) was simulated. The target was located at the range of $3$ -- $3.025\,\mathrm{km}$ away from the transceiver. The amplitudes and phases of the scattered signals from these closely spaced scatterers were random values with uniform distribution in $[0, 0.05]$ and uniform distribution in $[0, 2\pi]$, respectively. A beat signal with the SNR of $15\,\mathrm{dB}$ was synthesized by adding white Gaussian noise to consider measurement errors and thermal noise of the system and also contaminated by a strong interference with the same center frequency but a sweep slope of -0.98 times of that of the victim radar. The resultant beat signal is illustrated in Fig.~\ref{fig:disTar_SNR15_Sig_TD}\subref{fig:disTar_SNR15BeatSig_real}.    

Similar to the point target scenario, the interference-contaminated samples are first detected and cut out. The result is shown in the top panel in Fig.~\ref{fig:disTar_SNR15_Sig_TD}\subref{fig:disTar_SNR15_RawSig_real_mOd4}. Then, the signal model order was estimated by using the SAMOS method based on the other interference-free samples. However, due to the strong correlation among the beat signals scattered by the closely spaced scatterers, the model order was selected to be two by using the SAMOS method, which is significantly different from the theoretical value fifteen (see Fig.~\ref{fig:disTar_SNR15_Sig_TD}\subref{fig:disTar_SNR15_mOd}). So the SAMOS method cannot work properly in such scenarios. To investigate the reason of the failure of the SAMOS, we checked the singular value distribution of the matrix used for model order selection, as shown in Fig.~\ref{fig:disTar_SNR15_Sig_TD}\subref{fig:disTar_SNR_SV}. Based on Fig.~\ref{fig:disTar_SNR15_Sig_TD}\subref{fig:disTar_SNR_SV}, it is obvious that a proper model order should be not smaller than four. Taking the model order of four, the signal samples in the cut-out region were recovered by using the Burg-, IVM- and MP-based methods, which are shown in the two bottom plots in Fig.~\ref{fig:disTar_SNR15_Sig_TD}\subref{fig:disTar_SNR15_RawSig_real_mOd4} and Fig.~\ref{fig:disTar_SNR15_Sig_TD}\subref{fig:disTar_SNR15_RawSig_mOd4_IVM}, respectively. It is clear that the IVM-based interpolation is not stable and a blow-up is observed in Fig.~\ref{fig:disTar_SNR15_Sig_TD}\subref{fig:disTar_SNR15_RawSig_mOd4_IVM}. Meanwhile, compared to the Burg-based method, the proposed MP-based method reconstructed the signal samples with the best agreement with the reference signal (see the bottom plot in Fig.~\ref{fig:disTar_SNR15_Sig_TD}\subref{fig:disTar_SNR15_RawSig_real_mOd4}). Taking the FFT of the signal obtained after zeroing and the recovered signals with Burg- and MP-based methods, the related range profiles of targets were constructed and shown in Fig.~\ref{fig:disTar_SNR15_MP_Range_zoom_3km}\subref{fig:disTar_SNR15_MP_RP_zoom_3km_mOd4}. As expected, the range profile of the targets constructed with the signal recovered with the MP-based method has the best agreement with that formed using the reference signal. For quantitative evaluation, the RSNRs of the beat signals recovered with the Burg- and MP-based methods are obtained as $7.07\,\mathrm{dB}$ and $10.66\,\mathrm{dB}$, respectively. Their correlation coefficients relative to the reference signal are $0.8975e^{-0.0311}$ and $0.9584e^{0.0443}$. So the RSNRs and correlation coefficients confirm that the MP-based method gets more accurate signal reconstruction in the cut-out region than the Burg-based method.

Moreover, we also reconstructed the signal samples in the cut-out region using the three methods by setting the model order to be fifteen. Again, a blow-up as in Fig.~\ref{fig:disTar_SNR15_Sig_TD}\subref{fig:disTar_SNR15_RawSig_mOd4_IVM} is observed in the recovered signal by the IVM-based method (here the figure is omitted for conciseness). So it indicates that the instability of the IVM-based method may not be caused by the underestimation of the signal model order.  
Meanwhile, the recovered signal by the Burg-based method is still less accurate than that obtained with the MP-based method (see Fig.~\ref{fig:disTar_SNR15_Sig_TD}\subref{fig:disTar_SNR15_RawSig_real_mOd15} and Fig.~\ref{fig:disTar_SNR15_MP_Range_zoom_3km}\subref{fig:disTar_SNR15_MP_RP_zoom_3km_mOd15}). The RSNR and correlation coefficient of the recovered signal by the Burg-based method are $5.98\,\mathrm{dB}$ and $0.8779e^{0.0324}$ and their counterparts for the signal reconstructed with the MP-based method are $11.48\,\mathrm{dB}$ and $0.9663e^{0.0639}$, which further confirms that the MP-based method is superior to the Burg-based one in term of the signal reconstruction accuracy.

Finally, we want to mention that when multiple point targets in the same range bin are very close to each other, the Burg-based method could occasionally outperform the proposed MP-based method (for conciseness, we do not show it here). As the close targets in a range bin results in highly correlated beat frequencies, the characteristic polynomial of the corresponding AR model has many closely spaced roots.
The proposed MP-based method tends to estimate some dominant sinusoidal components (i.e., roots) that are close to the real roots in the mean square error sense while the Burg-based method attempts to estimate the coefficients of the characteristic polynomial of the AR model. Apparently, the latter operation is easier in such cases; thus, the Burg-based method results in more accurate signal estimation.  

\subsection{Effect of the length of interferences and SNR}

\begin{figure*}[!t]
    \centering \vspace{-5mm}
    \subfloat[]{
    \includegraphics[width=0.28\textwidth]{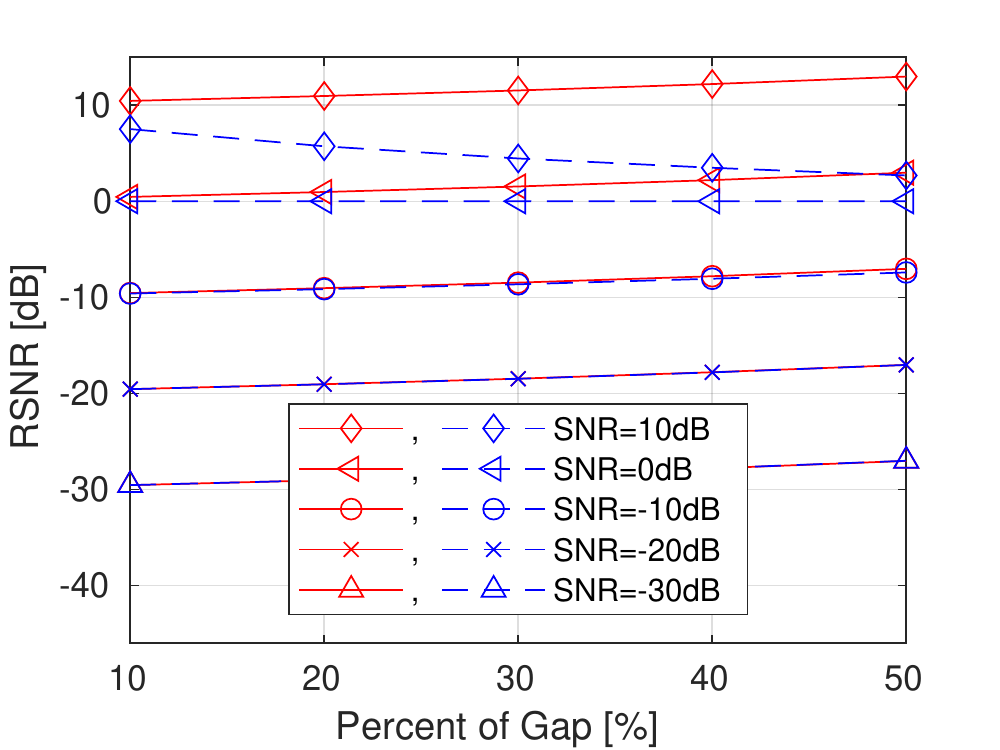}
    \label{fig:metric_RSNR_gap_MP}
    }
    \subfloat[]{
    \includegraphics[width=0.28\textwidth]{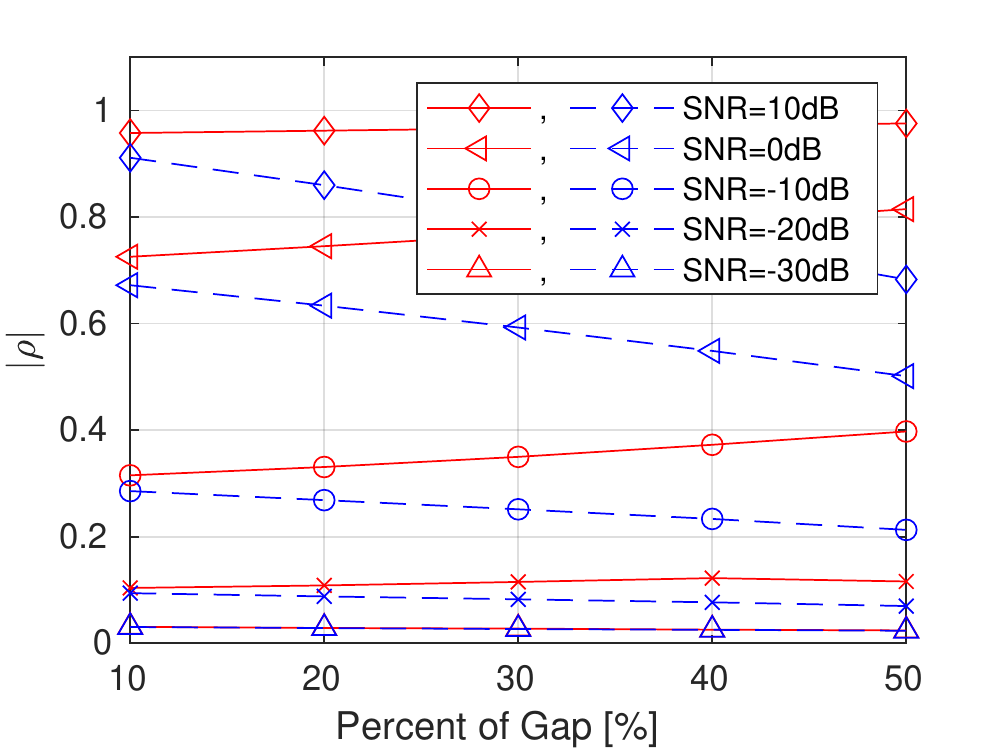}
    \label{fig:metric_Corr_gap_MP}
    }
    \subfloat[]{
    \includegraphics[width=0.28\textwidth]{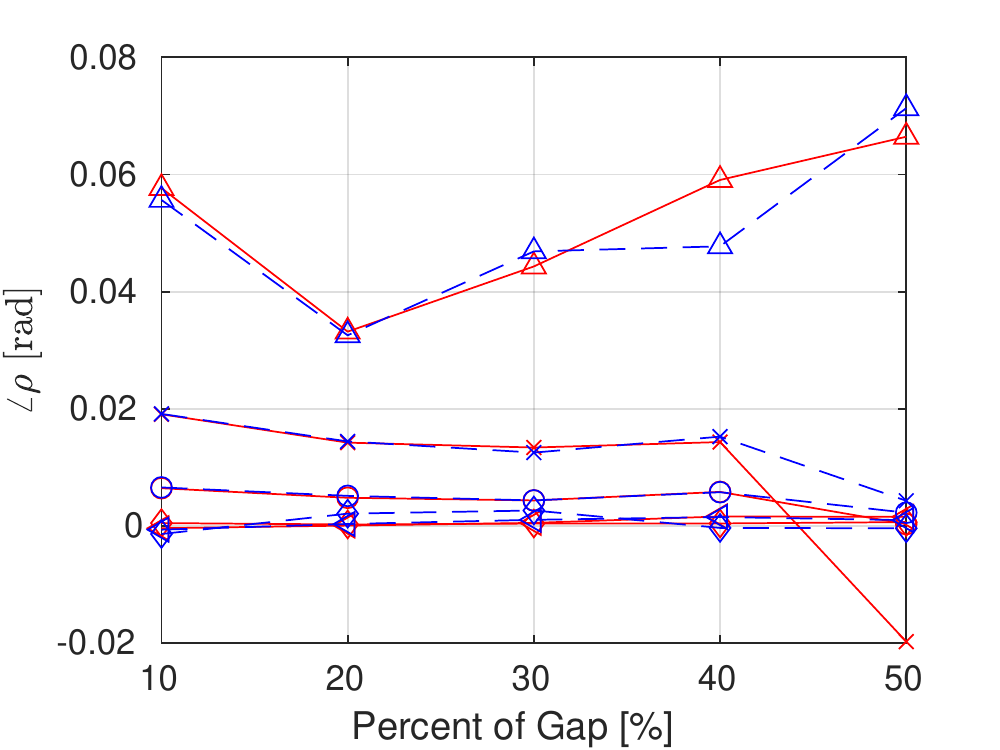}
    \label{fig:metric_Corr_Ang}
    }
    
    \caption{Impact of gap duration and SNR on the accuracy of the reconstructed signals with the Burg- (dashed blue lines) and MP-(solid red lines) based methods. \protect\subref{fig:metric_RSNR_gap_MP} shows the RSNRs with different gap durations. \protect\subref{fig:metric_Corr_gap_MP} and \protect\subref{fig:metric_Corr_Ang} show the moduli and phases of the correlation coefficients with respect to different gap durations.}
    \label{fig:metric_MP}
\end{figure*}

The impact of the interference duration (equivalently, the size of the cut-out gap caused by interference suppression) and the SNR on the performance of the proposed MP-based method for signal recovery is investigated in this section. For generality, the size of a cut-out gap is denoted by the ratio between the number of the removed interference-contaminated samples and the number of all signal samples in a sweep. The parameters for section~\ref{subsec:point_target_scenario} point target scenario simulation were used here. In the simulation, the SNR changes from $-30\,\mathrm{dB}$ to $10\,\mathrm{dB}$ with steps of $10\,\mathrm{dB}$ and at each SNR the interference duration increases from $10\%$ to $50\%$ with steps of $10\%$. To investigate the statistical performance of the proposed MP-based approach, 100 Monte Carlo runs were conducted at each SNR. The average RSNRs and correlation coefficients of the signals recovered by Burg- and MP-based methods are shown in Fig.~\ref{fig:metric_MP} (due to the blow-ups of signals recovered the IVM-based method, the corresponding RSNRs and correlation coefficients cannot be computed and are omitted here).     


From Fig.~\ref{fig:metric_MP}\subref{fig:metric_RSNR_gap_MP}, one can see that the RSNRs of the signals reconstructed with the Burg- and MP-based methods are almost identical when the SNR is smaller than $0\,\mathrm{dB}$. Meanwhile, they gradually improve and are larger than the SNRs with the increase of the size of the cut-out region. By contrast, when the SNR is equal to/larger than $0\,\mathrm{dB}$ the RSNRs of the signals obtained with the Burg- and MP-based methods show different changing trends (i.e., increase for MP-based method while keep steady/decrease for Burg-based method) with the widening of the cut-out gap. This is because that the cut-out operation eliminates not only the interference but also the noise in the interference-contaminated signal samples. When $\text{SNR}<0\,\mathrm{dB}$, the eliminated noise power is larger than that of the useful signals; thus, the RSNR would be larger than the SNR as long as the useful signal samples in the cut-out region can be recovered with certain accuracy with either Burg- and MP-based methods. However, when $\text{SNR}\geq 0\,\mathrm{dB}$, more signal power is suppressed than the noise power. The MP-based method jointly uses the signal samples at both sides of the gap to accurately recover the data in the cut-out region via an iterative scheme. The recovered signal could be equivalently regarded as the filtered samples, getting higher RSNR than the SNR of the original signal. In particular, when the cut-out gap occupies $50\%$ of the whole sweep, almost half of the noise power is suppressed; thus, $3\,\mathrm{dB}$ improvement of RSNR relative to the SNR of the input signal can be obtained as long as the useful signal samples in the cut-out region are accurately reconstructed (see Fig.~\ref{fig:metric_MP}\subref{fig:metric_RSNR_gap_MP}). On the other hand, the Burg-based method separately extrapolates the signal samples in the cut-out gap from both sides.  Its extrapolation accuracy degrades rapidly with the widening of the cut-out region, which causes larger signal difference (especially, large phase differences) between the reconstructed signal and the reference and thus makes its RSNR even worse than the original SNR. Therefore, in terms of the RSNR of the recovered signal, the Burg- and MP-based obtain comparable results when $\text{SNR}<0\,\mathrm{dB}$ while the latter one outperforms the former one when $\text{SNR}\geq 0\,\mathrm{dB}$.  

However, Fig.~\ref{fig:metric_MP}\subref{fig:metric_Corr_gap_MP} shows that the MP-based method constantly obtains comparable/better signal reconstruction compared to the Burg-based method regarding the modulus of the correlation coefficient. Moreover, with the increase of the SNR and the interference duration, the performance advantage of the MP-based method to the Burg-based one becomes larger. However, the phase of the correlation coefficient between the recovered signals with both methods and the reference are comparable when the interference duration is smaller than $40\%$ (Fig.~\ref{fig:metric_MP}\subref{fig:metric_Corr_Ang}). It gradually reduces to zero with the increase of the SNR of the original signal. Therefore, according to the above analyses, the MP-based method generally gets more accurate signal reconstruction than the Burg-based method in terms of both RSNR and correlation coefficient of the recovered signal.

\subsection{Computational Efficiency}
Both Burg- and IVM-based methods are very computational efficient as they just separately extrapolate the data in the gap from both sides. By contrast, the proposed MP-based method uses the SVD and an iterative scheme to jointly recover the signal in the cut-out region. So its computational load is slightly heavier than that of the Burg- and IVM based methods, which depends on the number of iterations in practice. For a scenario with moderate interference duration (20\%-30\%) and SNR, the MP-based method generally needs several iterations. Specifically, for the simulation in section~\ref{subsec:point_target_scenario}, it took $0.02\,\mathrm{s}$, $0.15\,\mathrm{s}$ and $27.05\,\mathrm{s}$ for the Burg-, IVM- and MP-based methods, respectively, when they were implemented in MATLAB and run on a computer with Intel Core i5-3470 Central Unit Processor (CPU) @ 3.2GHz and 8GB Random Access Memory (RAM). In this case, four iterations were executed in the MP-based method. To accelerate the MP-based method, Lanczos iteration \cite{golub2013matrix} or randomized algorithm \cite{Li2017Alg971} for the SVD could be exploited in future.

\section{Experimental results} \label{sec: Exp_Results}
In this section, experimental results with radar observations of an industrial chimney and raindrops are presented to demonstrate the effectiveness and accuracy of the proposed MP-based interference mitigation method.

\subsection{Experimental Setups}
The experiments used the TU Delft PARSAX \cite{Oleg2010} S-band (3.1315 GHz) radar system which is a full-polarimetric FMCW radar with two independent highly linear polarimetric RF channels in both transmitter and receiver. In the experiments, we consider the interference problem among the different polarimetric signals scattered from targets when the full-polarimetric radar simultaneously emits both horizontally and vertically polarized signals through the two transmitting channels and simultaneously acquires the scattered full-polarimetric signals. Specifically, we use the up- and down-chirp signals for simultaneous transmission on the horizontal (H-pol) and vertical (V-pol) polarization channels of the PARSAX radar, respectively. Then, the HV- (H-pol transmission, V-pol reception) and VV-(V-pol transmission, V-pol reception) polarimetric signals scattered from the same target would arrived at the V-pol receiving antenna at the same time. Although the up- and down-chirp waveforms are of great help to distinguish the scattered HV-pol and VV-pol signals, the strong VV-pol signal would still cause strong interference in the output of the HV-pol receiving channel. This kind of the interferences is categorized as Case 2 in Fig.~\ref{fig:interference_4_cases}. 

\begin{figure}[!t]
    \centering
    \vspace{-5mm}
    \subfloat[]{
    \includegraphics[width=0.22\textwidth]{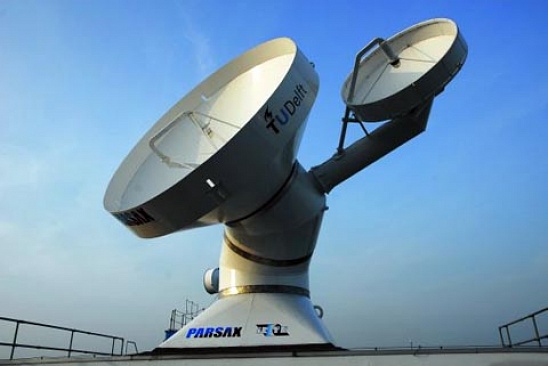}
    \label{fig:PARSAX}
    }
    \subfloat[]{
    \includegraphics[width=0.22\textwidth]{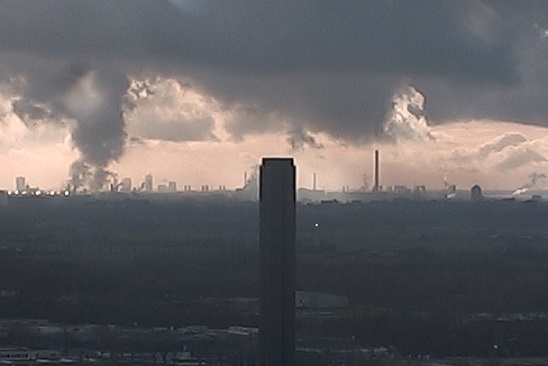}
    \label{fig:Chimney_camera}
    }
    \caption{Experimental measurement setup. \protect\subref{fig:PARSAX} shows PARSAX radar on the roof of EEMCS Faculty building and \protect\subref{fig:Chimney_camera} the industrial chimney used as a stationary target.}
    \label{fig:PARSAX_Chimney}
\end{figure}

In Experiment 1, we considered an industrial chimney as a stationary target and took measurements for a single sweep. The chimney is about $1.07\,\mathrm{km}$ away from PARSAX radar, which is installed on the roof of the building of the faculty of Electrical
Engineering, Mathematics \& Computer Science (EEMCS), TU Delft. The PARSAX radar is shown in Fig.~\ref{fig:PARSAX_Chimney}\subref{fig:PARSAX} and an image of the chimney captured by a camera with the same orientation as the radar is presented in Fig.~\ref{fig:PARSAX_Chimney}\subref{fig:Chimney_camera}. In Experiment 2, we observed a rain storm, which can be considered as a distributed target, by pointing the PARSAX radar vertically. The parameters for experimental measurements are listed in Table~\ref{Table: Experimental setup}.  

{ 
	\renewcommand{\arraystretch}{1.05}
\begin{table}[!t]
    \centering
    \caption{Experimental setup Parameters for Experiment 1 and Experiment 2}
    \label{Table: Experimental setup}
\begin{tabular}{l|l}
\hline\hline
\textbf{Parameter}          & \textbf{Value}  \\ \hline
Center frequency           & 3.1315 GHz                            \\ \hline
Bandwidth                           & 40           MHz                            \\ \hline
Time duration of a sweep                          & 1    ms                             \\ \hline
Number of samples per sweep                 & 16384                                    \\ \hline
Maximum range                       & 18.75    km                              \\ \hline
Number of sweeps per CPI  & 512  \\ \hline
Waveform                            & \begin{tabular}[c]{@{}l@{}l@{}}Simultaneous up- and  down-\\chirps on the H-pol and\\ V-pol polarization channels\end{tabular}  \\ \hline \hline
\end{tabular}
\end{table}
}

\subsection{Experiment 1: Stationary isolated target (Chimney)}

\begin{figure*}[!t]
    \centering 
    \vspace{-5mm}
    \subfloat[]{\hspace{-9mm}
    \includegraphics[width=0.7\textwidth]{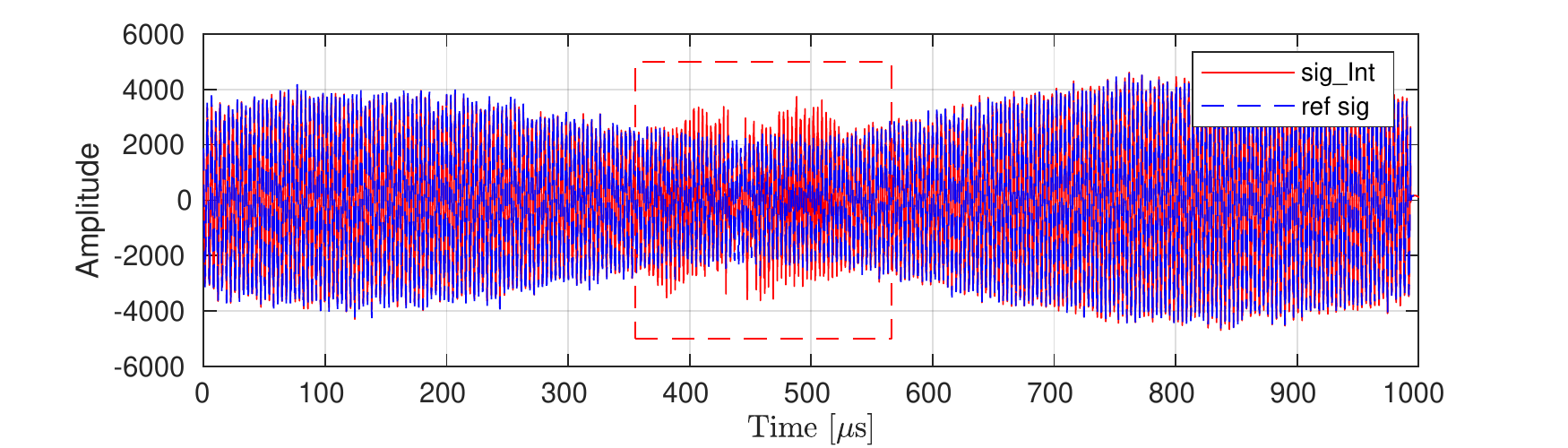}
    \label{fig:Chimney_RawData_IntRef}
    } 
    \subfloat[]{\hspace{-10mm}
    \includegraphics[width=0.265\textwidth]{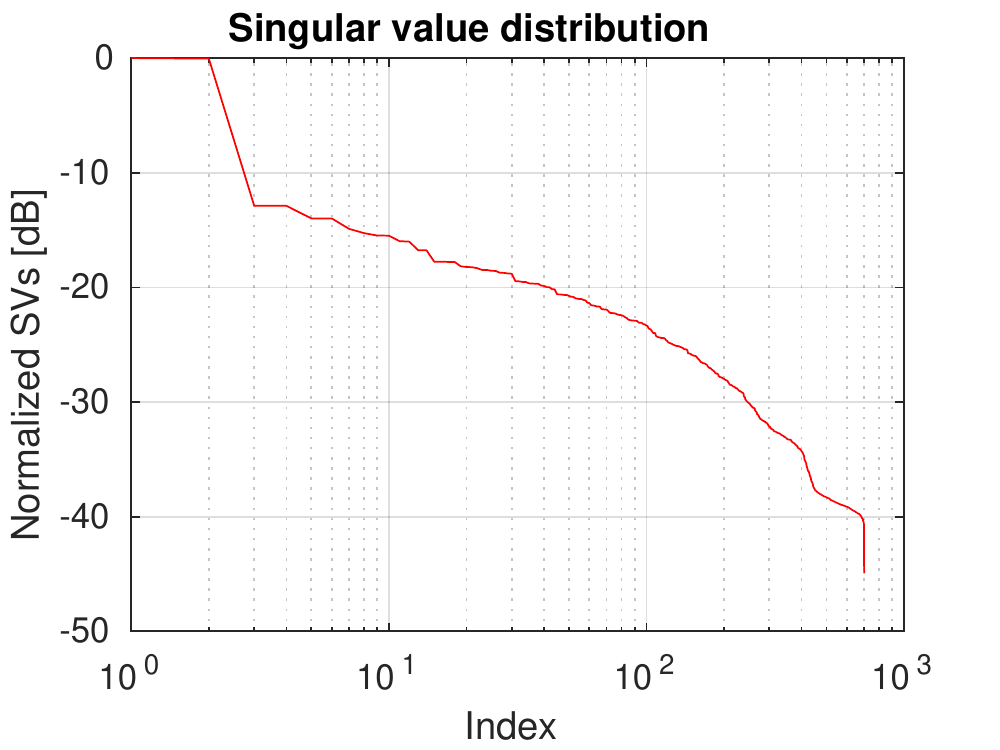}
    \label{fig:Chimney_SV}
    }    
    
    \vspace{-1mm}
    \subfloat[]{ \hspace{-10mm}
    \includegraphics[width=0.7\textwidth]{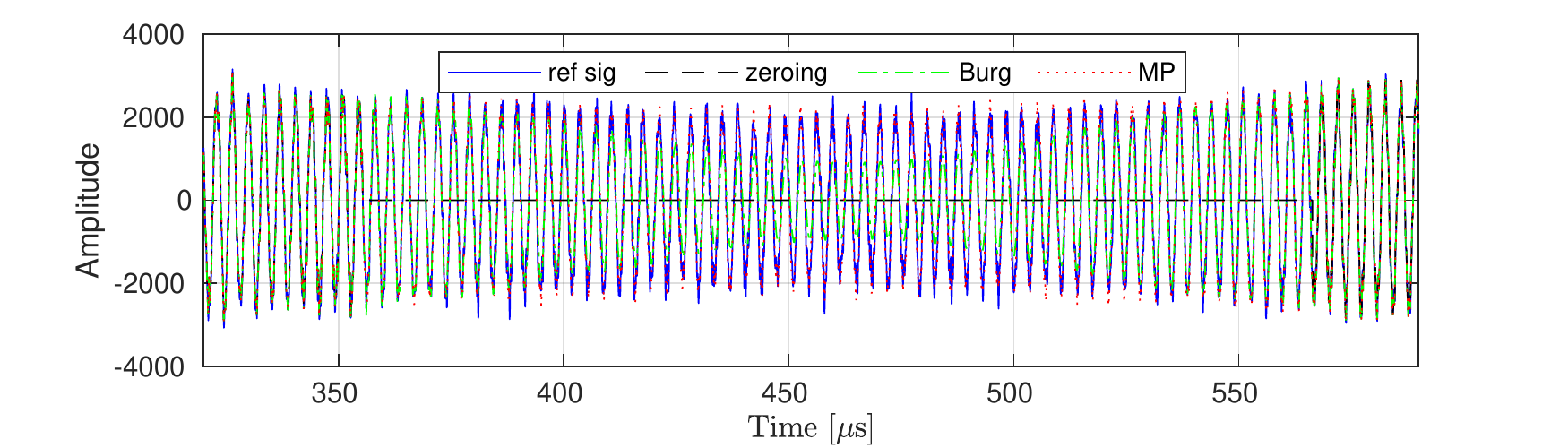}
    \label{fig:Chimney_RawData}
    }
    \subfloat[]{\hspace{-10mm}
    \includegraphics[width=0.265\textwidth]{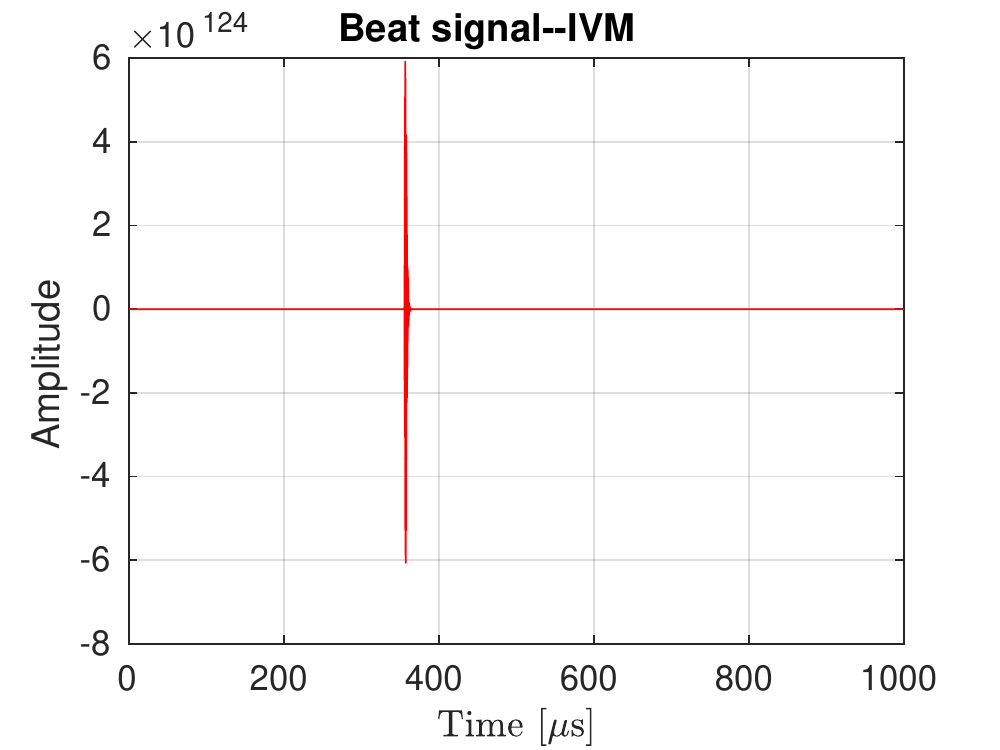}
    \label{fig:Chimney_rawData_IVM}
    }
    
    \caption{The beat signal acquired in one FMCW sweep for the chimney observation. \protect\subref{fig:Chimney_RawData_IntRef} shows the measured beat signals with (i.e., ``sig\_Int'' in the solid red line) and without the cross-polarimetric interference (i.e., ``ref sig'' in the dashed blue line). \protect\subref{fig:Chimney_RawData} presents the signals around the interference-contaminated region after interference mitigation using zeroing, Burg- and MP-based methods while \protect\subref{fig:Chimney_rawData_IVM} shows the recovered beat signal with the IVM-based method. }
    \label{fig:Exp_Chimney}
\end{figure*}

\begin{figure*}[!t]
    \centering 
    \vspace{-3mm}
    \subfloat[]{\hspace{-10mm}
    \includegraphics[width=0.6\textwidth]{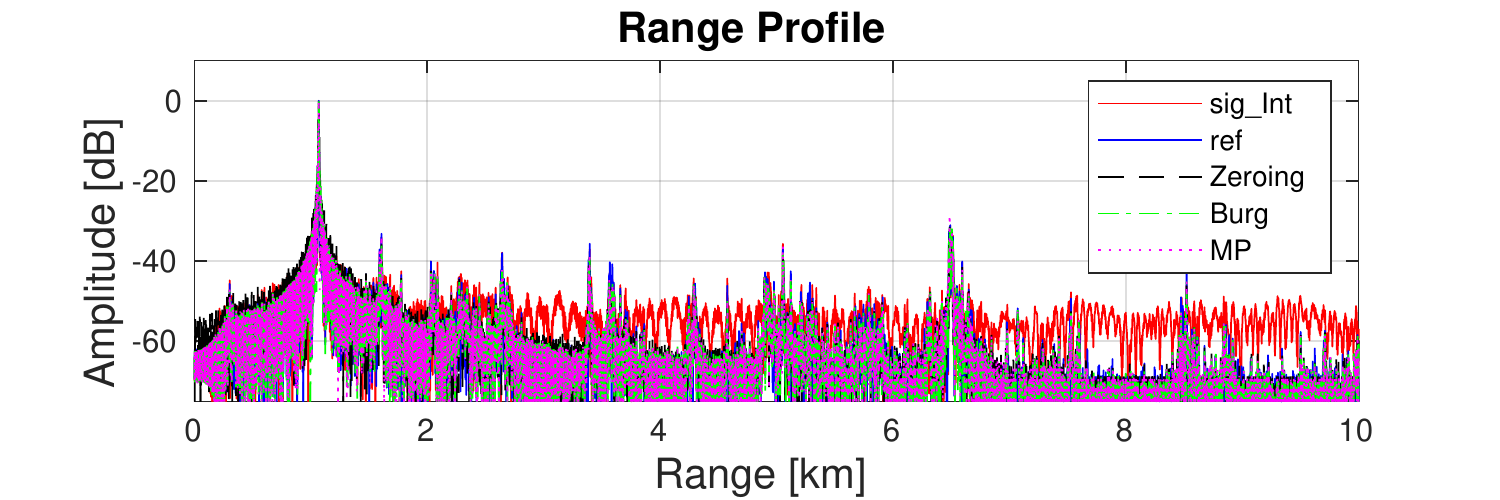}
    \label{fig:Chimney_RangeProfile_rec}
    }
    \subfloat[]{\hspace{-10mm}
    \includegraphics[width=0.25\textwidth]{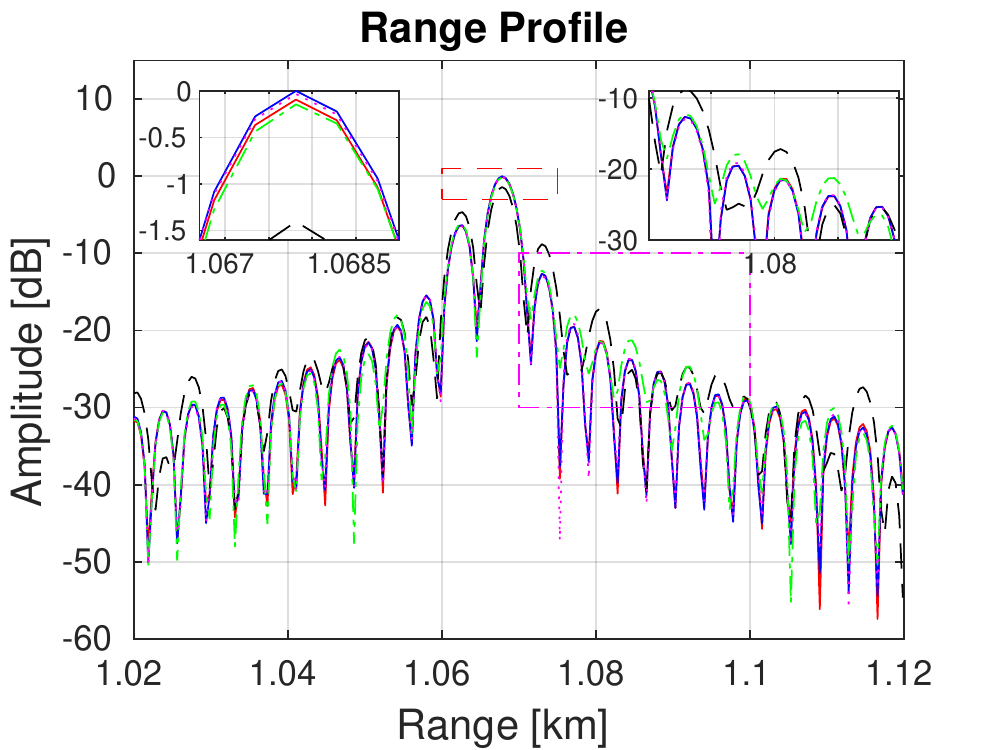}
    \label{fig:Chimney_RangeProfile_rec_1km}
    }
    \subfloat[]{\hspace{-5mm}
    \includegraphics[width=0.25\textwidth]{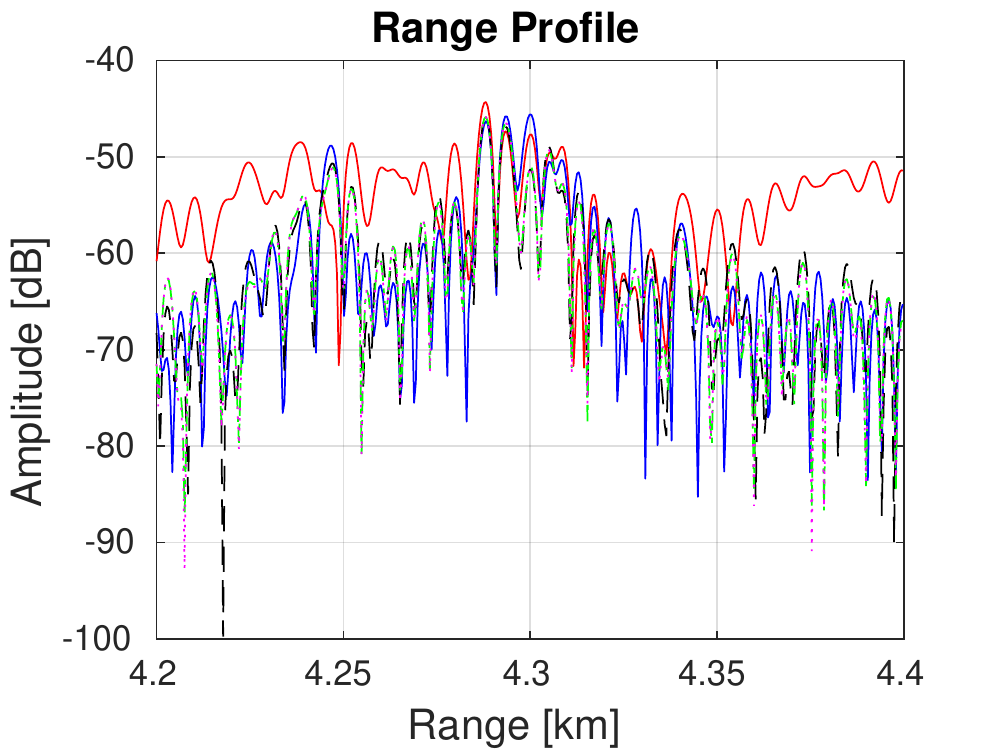}
    \label{fig:Chimney_RangeProfile_rec_4km}
    }
    \caption{The range profiles of the Chimney scenario obtained with the signals before and after interference mitigation. \protect\subref{fig:Chimney_RangeProfile_rec} shows the range profiles of the scenario within $10\,\mathrm{km}$ from the radar. \protect\subref{fig:Chimney_RangeProfile_rec_1km} and \protect\subref{fig:Chimney_RangeProfile_rec_4km} are the zoomed-in views of the range profiles of the targets at the distances of $1.07\,\mathrm{km}$ and $4.3\,\mathrm{km}$ from the radar, respectively.   }
    \label{fig:Exp_Chimney_RangeProfile}
\end{figure*}

Fig.~\ref{fig:Exp_Chimney}\subref{fig:Chimney_RawData_IntRef} shows the acquired HV-pol beat signal (i.e., ``sig\_Int'' in the solid red line) when the transmitter simultaneously emitted the up- and down-chirp signals with opposite chirp rates through the two transmitting channels with horizontal and vertical polarizations, respectively. The acquired HV-pol beat signal was polluted by the strong VV-pol signal arrived together at the receiving antenna, and the interference-contaminated samples are indicated by the dashed red rectangle in Fig.~\ref{fig:Exp_Chimney}\subref{fig:Chimney_RawData_IntRef}. For comparison, the reference HV-pol signal (i.e., ``ref sig'' in the dashed blue line) acquired by transmitting a single H-pol up-chirp signal is also presented. 

To suppress the VV-pol interference, the received signal was processed by using the zeroing, Burg-, IVM- and the proposed MP-based interference mitigation methods and the results are shown in Fig.~\ref{fig:Exp_Chimney}\subref{fig:Chimney_RawData} and \subref{fig:Chimney_rawData_IVM}. Comparing the signals obtained by all four interference mitigation methods with the reference signal, the MP-based method almost accurately reconstructs the clipped samples in the interference-contaminated region while the Burg-based method recovers these samples with underestimated amplitudes. By contrast, the IVM-based method leads to a blow-up in the recovered beat signal (Fig.~\ref{fig:Exp_Chimney}\subref{fig:Chimney_rawData_IVM}), which again shows its instability. In addition, before applying the Burg-, IVM- and MP-based methods to recover the signal samples in the cut-out region, SAMOS was used to estimate the signal model order and a model order of two was selected, which is highly underestimated considering the complex environment surrounding the chimney. Hence, we decided to select the model order empirically based on the normalized singular value distribution of the matrix used by SAMOS (see Fig.~\ref{fig:Exp_Chimney}\subref{fig:Chimney_SV}). With a threshold of $10^{-2}$ (i.e., $20\,\mathrm{dB}$) for the normalized SVs, a model order of 40 was selected and used by the three methods for signal reconstruction.

Moreover, the range profiles constructed with the interference-contaminated signal, reference signal, the signals acquired after interference mitigation are displayed in Fig.~\ref{fig:Exp_Chimney_RangeProfile}\subref{fig:Chimney_RangeProfile_rec} (due to invalid signal recovery of the IVM-based method, its RP is omitted). It is clear that the range profile obtained with the interference-contaminated signal has higher ``noise floor'' in contrast to that formed with other signals, which would mask weak targets. For the convenience of comparison, the close-ups of the range profiles of the chimney at the distance of $1.07\,\mathrm{km}$ and some weak targets at the distance of  $4.3\,\mathrm{km}$ in Fig.~\ref{fig:Exp_Chimney_RangeProfile}\subref{fig:Chimney_RangeProfile_rec} are shown in Fig.~\ref{fig:Exp_Chimney_RangeProfile}\subref{fig:Chimney_RangeProfile_rec_1km} and \subref{fig:Chimney_RangeProfile_rec_4km}. From Fig.~\ref{fig:Exp_Chimney_RangeProfile}\subref{fig:Chimney_RangeProfile_rec_4km}, a clear peak for a weak target at the distance of $4.24\,\mathrm{km}$ can be observed in the range profiles generated with the reference signals and the signals acquired after interference mitigation. By contrast, a deep null is seen at the same position in the range profile formed with the interference-contaminated signal, which could be caused by the destructive interference between the interference and the target's signal. Moreover, the range profiles obtained with signals after mitigating the interference by using Burg- and MP-based methods are comparable to the reference one and have lower sidelobes for the weak targets around the distance of $4.3\,\mathrm{km}$. On the other hand, the range profile of the chimney acquired after processing with the proposed MP-based interference mitigation is almost identical to the one formed with the reference signal. However, the zeroing caused a void of signal samples and the Burg-based method underestimated signal amplitude in the cut-out region; thus, they cause higher sidelobes and power loss in the constructed range profiles (see the insets in Fig.~\ref{fig:Exp_Chimney_RangeProfile}\subref{fig:Chimney_RangeProfile_rec_1km}).

\subsection{{Experiment 2: Distributed target (Rain)}} 

\begin{figure}[!t]
    \centering
    \includegraphics[width=0.36\textwidth]{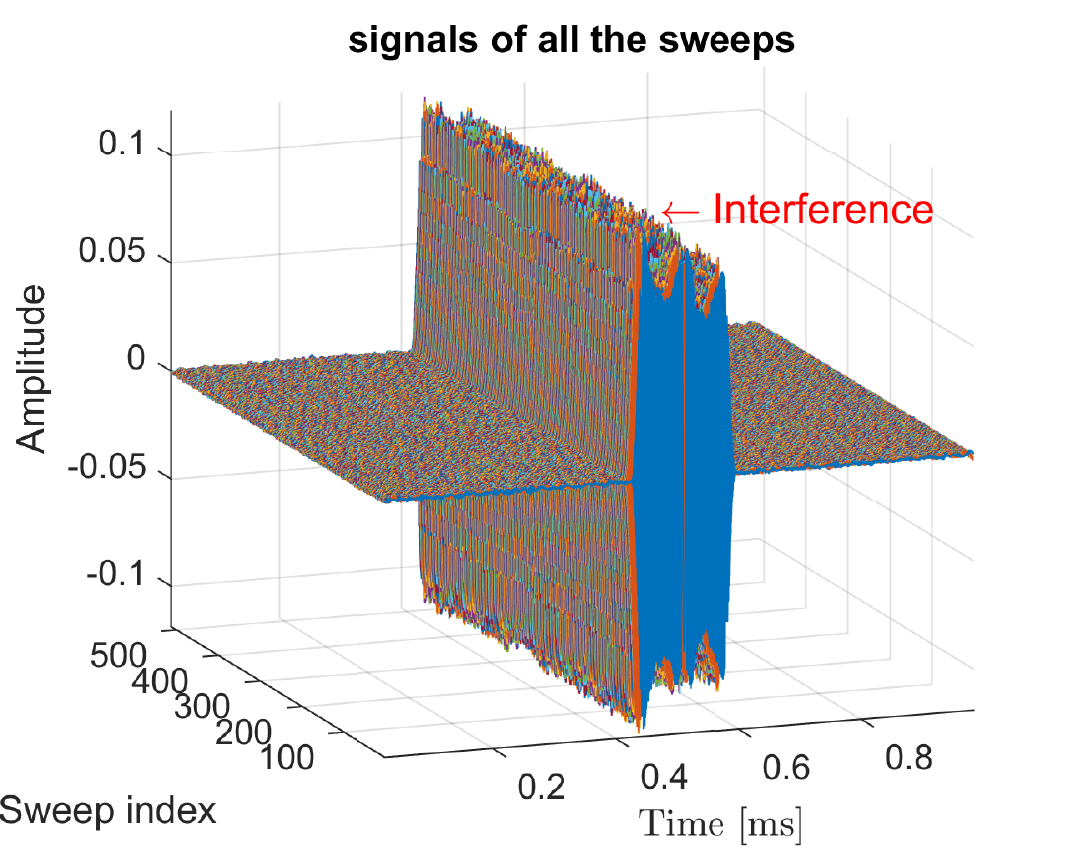}
    \caption{The signals of all the sweeps scattered from rain droplets.}
    \label{fig:rain_sig_AllSweep}
\end{figure}

\begin{figure}[!t]
    \centering
    \vspace{-3mm}
    \subfloat[]{
    \includegraphics[width=0.42\textwidth]{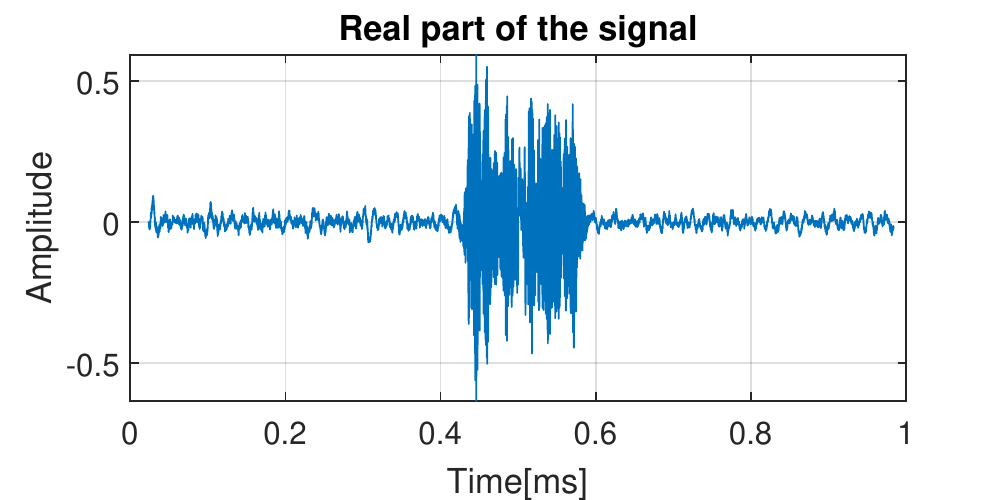}
    \label{fig:rain_sig_sweep_intf_AftDopFT}
    }
    
    \vspace{-4mm}
    \subfloat[]{
    \includegraphics[width=0.42\textwidth]{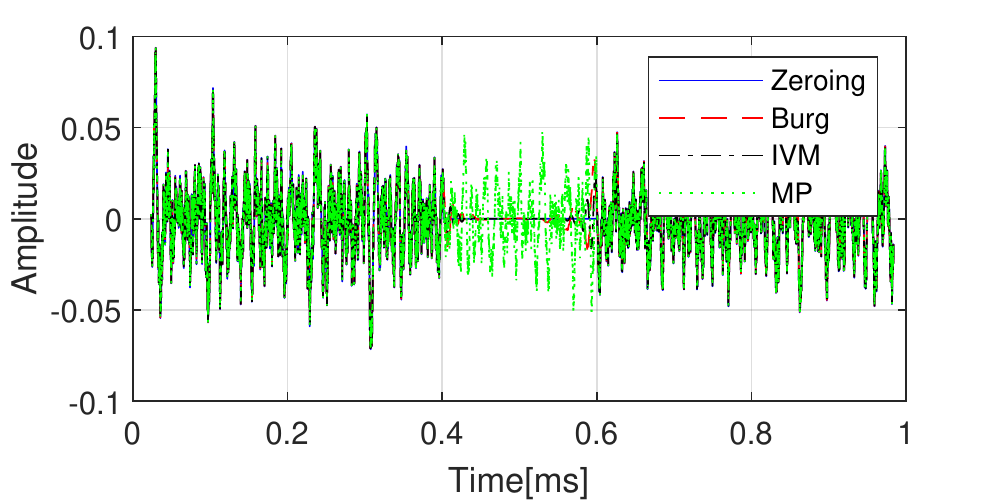}
    \label{fig:rain_sig_sweep_recons}
    }
    \caption{The time signals at a Doppler bin after taking FFT along the slow-time dimension. \protect\subref{fig:rain_sig_sweep_intf_AftDopFT} and \protect\subref{fig:rain_sig_sweep_recons} show the time signal before and after interference mitigation. }
    \label{fig:Exp_rain_timeSig}
\end{figure}

\begin{figure*}[!t]
    \centering
    \vspace{-3mm}
    \subfloat[]{
    \includegraphics[width=0.24\textwidth]{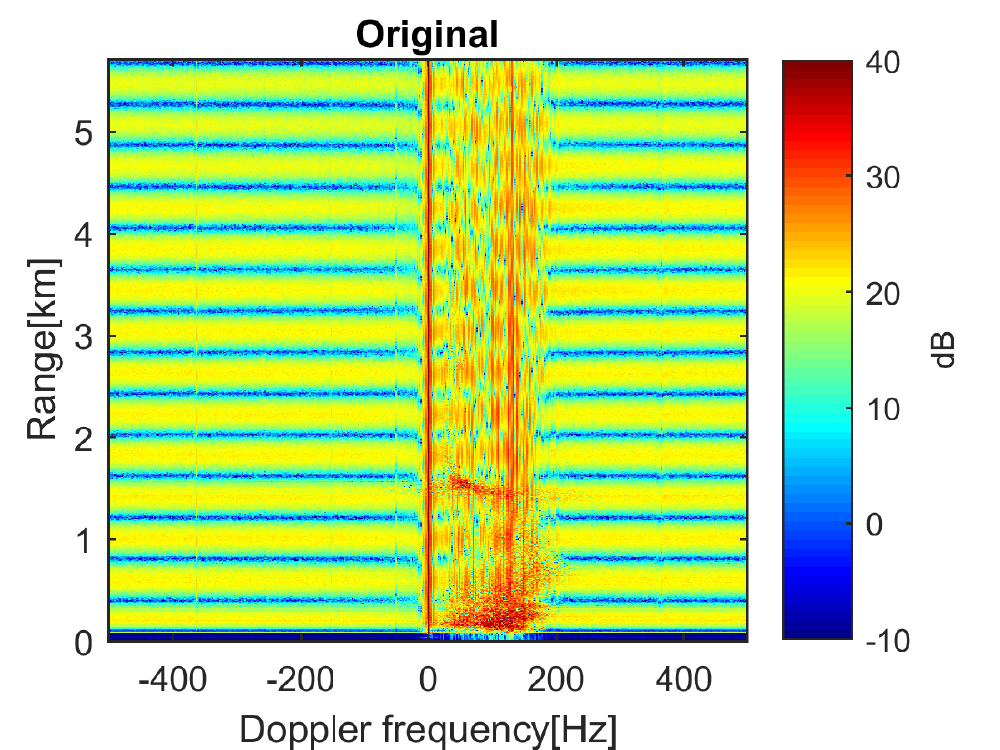}
    \label{fig:rain_RD_orig}
    }
    \subfloat[]{
    \includegraphics[width=0.24\textwidth]{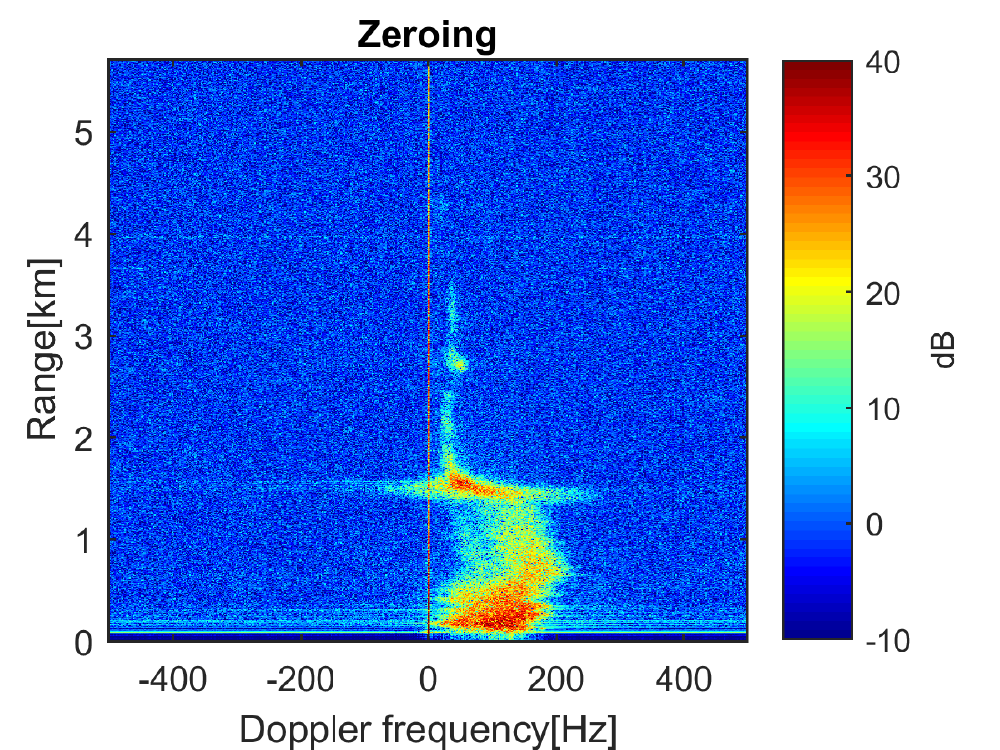}
    \label{fig:rain_RD_Zeroing}
    }
    \subfloat[]{
    \includegraphics[width=0.24\textwidth]{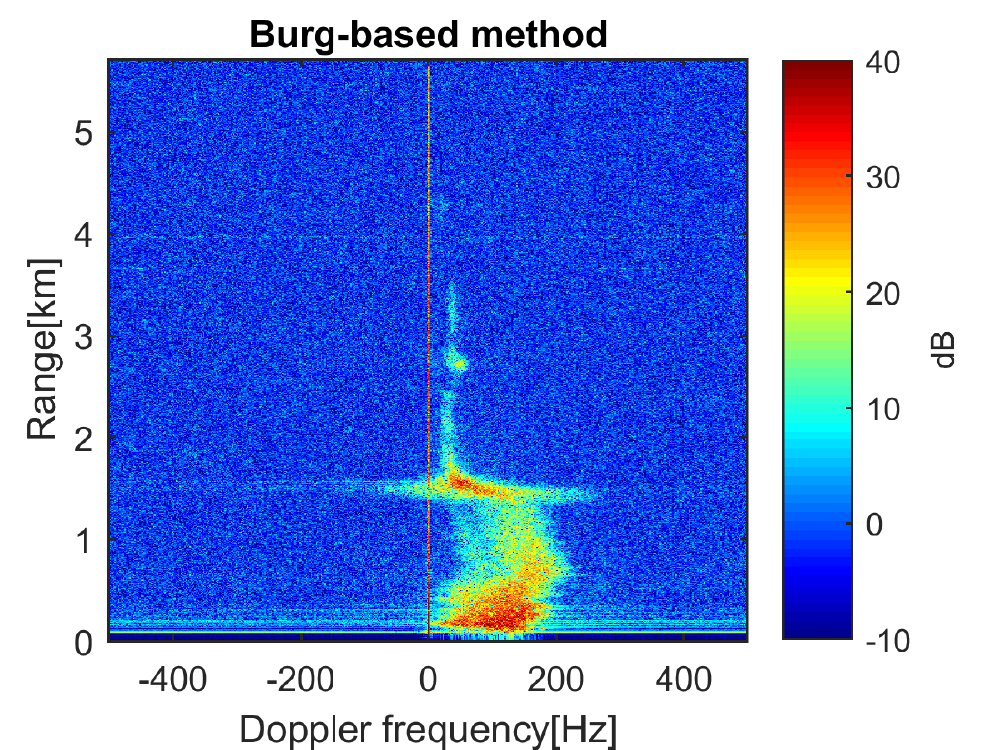}
    \label{fig:rain_RD_burg}
    }
    \subfloat[]{
    \includegraphics[width=0.24\textwidth]{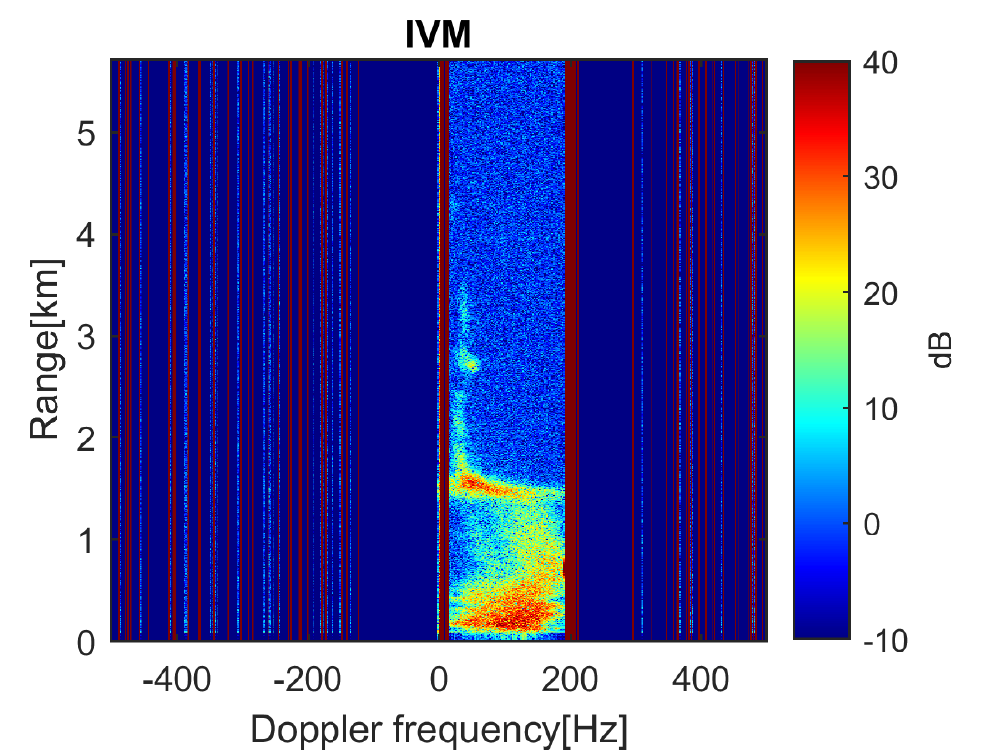}
    \label{fig:rain_RD_IVM}
    }
    
    \subfloat[]{
    \includegraphics[width=0.24\textwidth]{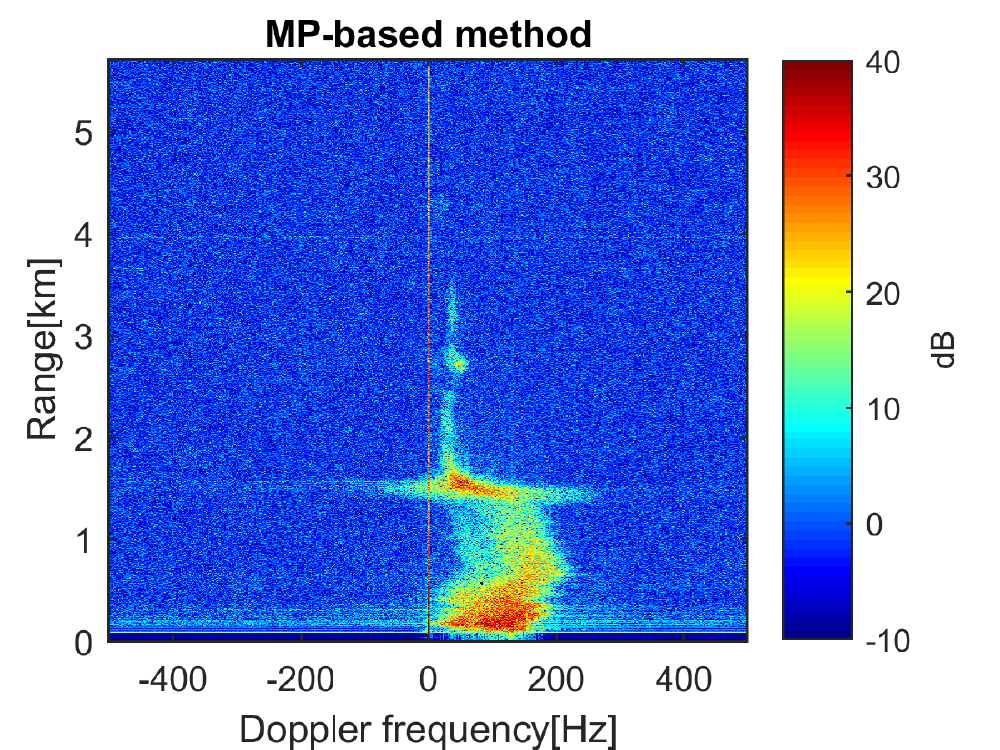}
    \label{fig:rain_RD_MP}
    }
    \subfloat[]{
    \includegraphics[width=0.24\textwidth]{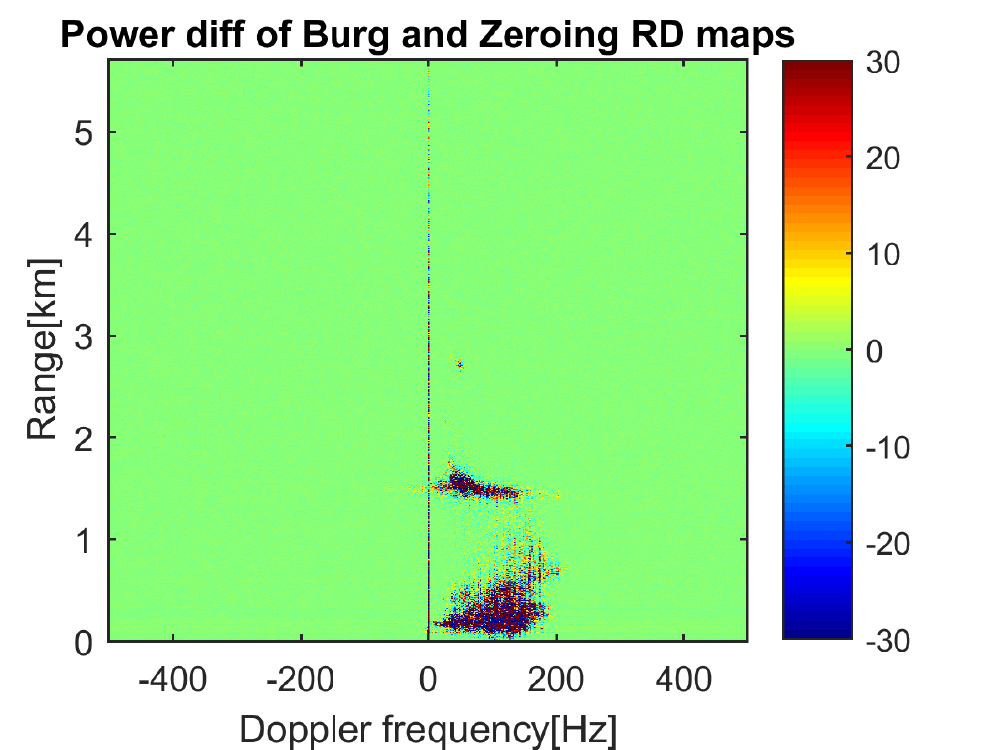}
    \label{fig:rain_RD_Diff_burg_Zero}
    }
    \subfloat[]{
    \includegraphics[width=0.24\textwidth]{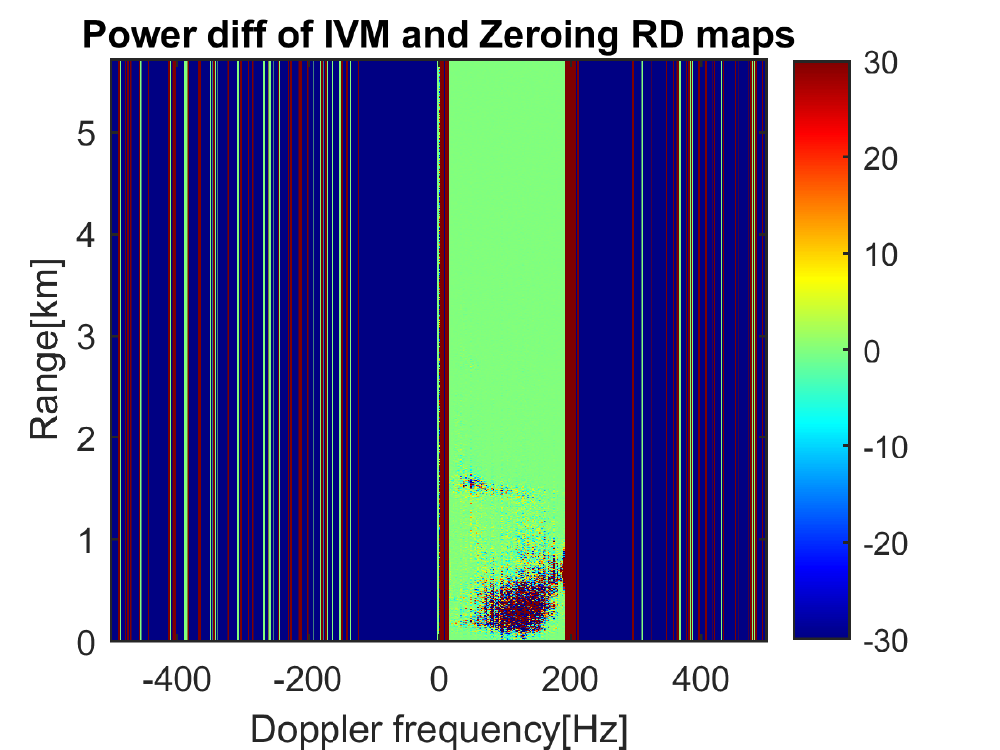}
    \label{fig:rain_RD_Diff_IVM_Zero}
    }
    \subfloat[]{
    \includegraphics[width=0.24\textwidth]{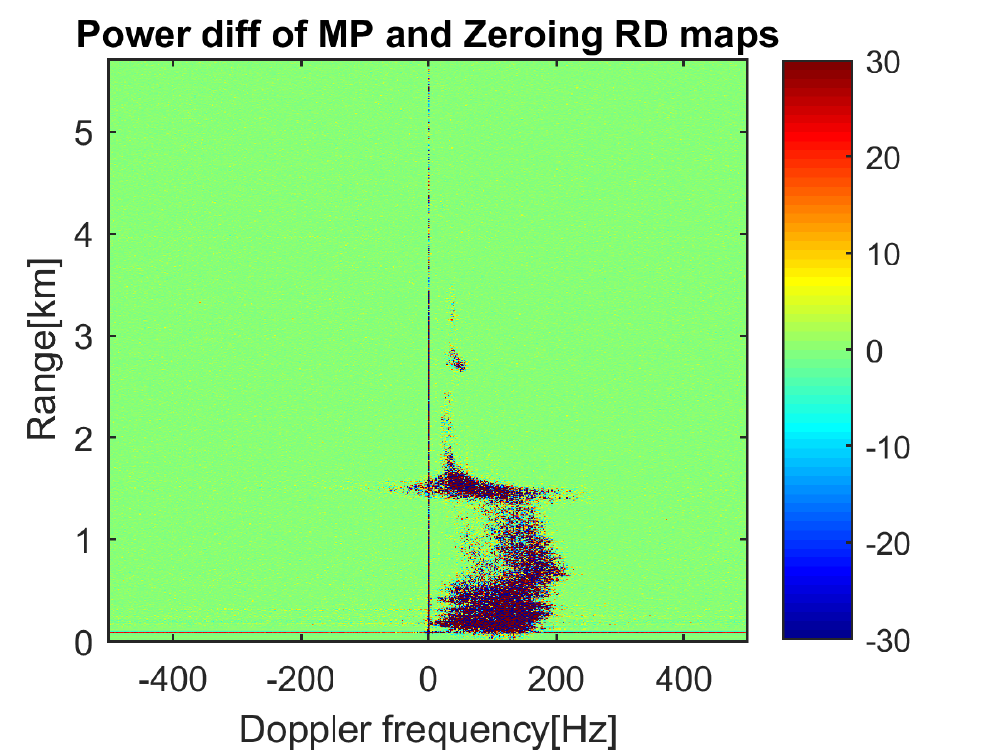}
    \label{fig:rain_RD_Diff_MP_Zero}
    }
    
    \caption{The range-Doppler processing results of the rain data. \protect\subref{fig:rain_RD_orig} is the RD map obtained with the original interference-contaminated signals. \protect\subref{fig:rain_RD_Zeroing}, \protect\subref{fig:rain_RD_burg}, \protect\subref{fig:rain_RD_IVM}, and \protect\subref{fig:rain_RD_MP} are formed by the signals after interference mitigation by using the zeroing, Burg-, IVM- and MP-based methods, respectively. \protect\subref{fig:rain_RD_Diff_burg_Zero}, \protect\subref{fig:rain_RD_Diff_IVM_Zero}, and \protect\subref{fig:rain_RD_Diff_MP_Zero} show the corresponding power differences between the RD maps in \protect\subref{fig:rain_RD_burg}-\protect\subref{fig:rain_RD_MP} and \protect\subref{fig:rain_RD_Zeroing}. }
    \label{fig:rain_RD}
\end{figure*}

In this experiment, we used 512 sweeps as a Coherent Processing Interval (CPI) for full-polarimetric measurements of rain droplets. After simple preprocessing to suppress the direct coupling, the acquired HV-pol signals in all the sweeps are shown in Fig.~\ref{fig:rain_sig_AllSweep}, where the interference-contaminated samples are located in the time interval from $0.4\,\mathrm{ms}$ to $0.6\,\mathrm{ms}$. The interference was caused by the VV-pol signals, which are generally much stronger than the desired HV-pol signals (see the much larger amplitudes of the interference-contaminated samples relative the rest ones). So after the range-Doppler (R-D) processing, the formed R-D map of the rain droplets is completely overwhelmed by the interference, as shown in Fig.~\ref{fig:rain_RD}\subref{fig:rain_RD_orig}.    

As the raindrops are moving targets, we suggest first taking the FFT with respect to the slow time in a CPI and then performing the interference mitigation to the time signal along each Doppler bin to avoid the possible detrimental impact of errors caused by interference mitigation on the Doppler information. Fig.~\ref{fig:Exp_rain_timeSig}\subref{fig:rain_sig_sweep_intf_AftDopFT} shows the time signal in a Doppler bin after taking the FFT along the slow time and the interference is still observed in the interval from $0.4\,\mathrm{ms}$ to $0.6\,\mathrm{ms}$. Applying the proposed MP-based interference mitigation method, zeroing, Burg- and IVM-based methods to this time signal, the resultant signals are presented in Fig.~\ref{fig:Exp_rain_timeSig}\subref{fig:rain_sig_sweep_recons}. The MP-based method successfully recovers the missing signals in the gap resulting from interference suppression while the Burg- and IVM-based methods reconstruct only the missing samples which are close to the front and rear available measurements with underestimated amplitudes. Note that for the rain data set, the SAMOS method could not estimate proper model orders, either. So we empirically determine the model order of the signal in each Doppler bin based on the normalized singular value distribution of the matrix used by SAMOS with a threshold of $10^{-4}$. The estimated signal model order was used by the Burg-, IVM- and MP-based methods to reconstruct the signal in the cut-out region. 

After mitigating the interferences for the time signals in all Doppler bins, an FFT is taken along the fast time to get the R-D map of the rain drops. Fig.~\ref{fig:rain_RD}\subref{fig:rain_RD_Zeroing}-\subref{fig:rain_RD_MP} present the obtained R-D maps in logarithmic scale of the moduli of signals after {interference mitigation} with zeroing, the Burg-, IVM- and MP-based methods, respectively. Except the R-D maps obtained with the IVM-based method, the other three R-D maps are visually almost identical and their qualities are noticeably improved compared to that obtained without interference mitigation (Fig.~\ref{fig:rain_RD}\subref{fig:rain_RD_orig}). 

Due to the lack of ground truth reference, we alternatively assess the improvement of the R-D maps obtained with the Burg-, IVM- and MP-based methods relative to the one got with zeroing by computing the power differences between the pixels of the R-D maps of the three signal reconstruction methods and zeroing method. The results are shown in linear scale in Fig.~\ref{fig:rain_RD}\subref{fig:rain_RD_Diff_burg_Zero}-\subref{fig:rain_RD_Diff_MP_Zero}. One can see that the power difference between the R-D maps of MP-based method and zeroing in Fig.~\ref{fig:rain_RD}\subref{fig:rain_RD_Diff_MP_Zero}, compared with that in Fig.~\ref{fig:rain_RD}\subref{fig:rain_RD_Diff_burg_Zero}, presents a pattern much closer to the R-D maps in Fig.~\ref{fig:rain_RD}\subref{fig:rain_RD_Zeroing}, \subref{fig:rain_RD_burg}, and \subref{fig:rain_RD_MP}. As in the rain data set the strong VV-pol interferences appear at the similar time interval in all the sweeps within the CPI, the zeroing method eliminates the signal samples within this time interval (i.e., between about $0.4\,\mathrm{ms}$ to $0.6\,\mathrm{ms}$) in all the sweeps. So the power difference of the R-D maps of zeroing and the other three methods are determined by the contribution of the beat signal samples in the cut-out region. Theoretically, the beat signals of rain droplets in the cut-out time interval in a CPI can be considered as the acquired data by using an FMCW radar with narrower bandwidth (i.e., shorter FMCW sweep duration) but keep other system parameters unchanged; thus, they can form a similar R-D map as that constructed with the full-sweep signals in the CPI but with lower range resolution. Namely, the more accurate the signal samples recovered by the Burg-, IVM- and MP-based methods in the cut-out region are, the closer to the actual R-D map the pattern of the power difference between the R-D maps of these methods and the zeroing approach. Therefore, the MP-based method gets more accurate estimation of the signals in the cut-out region than the Burg-based method. Furthermore, large portions of the positive power difference in Fig.~\ref{fig:rain_RD}\subref{fig:rain_RD_Diff_burg_Zero} and \subref{fig:rain_RD_Diff_MP_Zero} reveal that compared to the zeroing technique, both Burg- and MP based method improve the signal powers by reconstructing the missing signals in the cut-out region. In addition, due to the instability of the IVM-based method, the blow-ups in its reconstructed signals cause the streaks with very large amplitudes in many Doppler bins (Fig.~\ref{fig:rain_RD}\subref{fig:rain_RD_IVM}). So the accuracy of the recovered signals by the IVM-based method is worse than that of the Burg- and MP-based methods.             

{\color{red} }

\section{Conclusion} \label{sec: Conclusion}

In this paper, we present a matrix-pencil based interference mitigation method for FMCW radar systems. The proposed method exploits the feature of the desired beat signals as a sum of exponential sinusoidal components, which is different from the chirp-like waveforms of interferences after dechirping on reception, for interference suppression. The method is implemented in two steps by first detecting and cutting out the interference-contaminated samples and then recovering the signal samples in the cut-out region based on the exponential sinusoidal model of desired beat signals. It addresses the discontinuity of the signals caused by traditional zeroing technique and overcomes the power loss of useful signals. Meanwhile, it results in lower sidelobes of the range profile of a target. Moreover, compared to the Burg-based method, it significantly improves the accuracy of the estimated signals in the cut-out region by an iterative estimation scheme, which has demonstrated through both numerical simulations and experimental results. The numerical simulations also reveal that the proposed method can robustly work in the scenarios with a low signal to noise ratio (down to 0dB) and with a long interference duration (up to 50\% of a sweep). In addition, the proposed MP-based method can be extended to 2D or high-dimensional cases to mitigate interferences directly in a higher dimensional space (e.g., RD or range-DOA domains), especially for point-target scenarios, which would be considered in future work.


%

\section*{Acknowledgment}
The authors acknowledge the contribution of N. Cancrinus to this research by testing applicability of the matrix pencil method to the beat signal reconstruction.




\bibliographystyle{IEEEtran}
%
\bibliography{reference}

%








\end{document}